\documentclass[manuscript,screen,nonacm]{acmart}

\setcopyright{none}
\renewcommand\footnotetextcopyrightpermission[1]{}

\usepackage[nointegrals]{wasysym}   
\usepackage{colortbl} 
\usepackage{url}
\usepackage{longtable}
\usepackage{ragged2e}
\usepackage{hyperref}
\usepackage{lipsum}
\usepackage{enumitem}

\AtBeginDocument{%
  }
\usepackage{xspace}
\usepackage{booktabs}
\usepackage{tabularx}
\usepackage{float}

\definecolor{lightblue}{RGB}{212, 235, 255}
\definecolor{orange}{RGB}{255, 105, 0}
\definecolor{lightgreen}{RGB}{177, 231, 171}
\definecolor{lightyellow}{RGB}{255, 255, 148}
\definecolor{darkgreen}{RGB}{50, 200, 50}
\definecolor{darkyellow}{RGB}{230, 200, 0}

\newcolumntype{P}[1]{>{\centering\arraybackslash}p{#1}}
\newcolumntype{L}[1]{>{\raggedright\let\newline\\\arraybackslash\hspace{0pt}}m{#1}}
\newcolumntype{R}[1]{>{\raggedleft\arraybackslash}p{#1}}

\begin{document}

\title{U.S. State Guidance on AI in K-12 Education: A Comparative Content and Citation Analysis}

\author{Shahan Ali Memon}
\authornote{Authors contributed equally to this research.}
\email{samemon@uw.edu}
\orcid{0000-0002-1152-0867}
\author{Anna-Maria Gueorguieva}
\authornotemark[1]
\email{agueorg@uw.edu}
\orcid{0009-0006-4850-9318}
\author{Anukriti Kumar}
\email{anukumar@uw.edu}
\orcid{0000-0001-9441-3399}
\author{Lucy Lu Wang}
\email{lucylw@uw.edu}
\orcid{0000-0001-8752-6635}
\author{Jevin D. West}
\email{jevinw@uw.edu}
\orcid{0000-0002-4118-0322}
\author{Nic Weber}
\email{nmweber@uw.edu}
\orcid{0000-0002-3527-5974}
\affiliation{%
  \institution{Information School, University of Washington}
  \city{Seattle}
  \state{Washington}
  \country{USA}
}

\renewcommand{\shortauthors}{Memon et al.}

\begin{abstract}
In the years following the public release of ChatGPT in late 2022, more than half of U.S. states have published formal guidelines around artificial intelligence (AI) usage in K-12 education, creating a record of largely independent state education departments responding to a shared technological disruption. In this work, we analyze publicly available AI usage guidance documents from 35 states in the U.S.~and Puerto Rico. We compare the guidance in these documents across various dimensions, including pedagogy and instruction, audience, actionability, workforce and career readiness, responsible use and equity, and alignment with state-level demographics and administrative capacity. Finally, we investigate who authored these documents and their cited sources, identifying epistemic foundations underlying their claims. We find that guidance priorities converge on topics anchored in laws such as data privacy and academic integrity, each addressed in 97\% of documents. While most documents mention equity of access (86\%) and algorithmic bias (97\%), fewer discuss specific marginalized populations, such as rural or tribal communities (42\% and 14\%). Across documents, there is a common framing of urgency around AI integration to prepare students for workforce readiness (86\%) and promote loosely-defined \emph{personalized learning} (86\%); this urgency framing to use AI is not matched by specific guidance for capacity-building, accountability, and actionability. State priorities show no relationship to state demographics, capacity, or partisanship. Instead, we identify notable similarities and interdependencies between states with shared citations to non-profit organizations like TeachAI (78\%) and Code.org (44\%), with some states reproducing TeachAI's model guidance near-verbatim, and peer citations concentrating on a few early states such as Oregon and Washington. No state includes parents as contributors on these documents, and only five include students. Based on our findings, we provide a set of actionable recommendations for policymakers and practitioners.
\end{abstract}

\keywords{education, AI governance, policy, state guidance, policy diffusion}

\maketitle

\section{Introduction}
From the printing press and calculators to the internet, novel technologies reshape pedagogy and education norms. Artificial Intelligence (AI) is another such technology, the usage of which has drawn considerable attention and concern~\cite{yu2025exploring, dwyer-2026, al-marzouqi-2024}. Like other emerging technologies, the rapid advancement of AI has faced a regulatory lag~\cite{currie2025rethinking, bengio2024managing, marchant2011growing}. This problem is further compounded by the divergent attitudes surrounding the technology~\cite{o2023governs}, and its general-purpose nature which enables deployment across diverse downstream applications and complicates application-specific regulatory approaches~\cite{bommasani2021opportunities, wang2026distinguishing}.

AI governance is particularly salient in K-12 education. Decisions can influence foundational elements of teaching, curriculum priorities, upskilling, assessment, and access to learning opportunities, resources, and jobs. AI usage in classrooms has also outpaced the development of norms and guidelines, creating a notable gap in governance: A 2026 survey~\cite{zimmermann2026teens} conducted by Common Sense Media of 1,017 teenagers representing 50 states found that 70\% of the teens reported using AI tools for schoolwork. However, 73\% reported lack of key AI literacy conversations with their teachers, and 35\% reported not understanding their schools' rules around AI use. In 2026 Walton Family Foundation-Gallup surveys of K-12 educators, 59\% of the educators reported using AI for school work and teaching~\cite{brenan2026teachers}. However only 18\% reported receiving any formal guidance from school administrators~\cite{ash2026teachers}.

There are multiple policy instruments that can constrain or encourage specific behavior in school administration, by teachers, and by students, including passing state legislation, AI guidance documents from a state's Department of Education, Common Core curriculum, and district and school level policies. Among these, \emph{guidance} documents---most relevant to this study---play an important role in governance, training, and regulation. 
Unlike notice-and-comment style legislative rules that carry an enforceable \emph{force of law}, guidance documents provide nonbinding recommendations and are typically issued by agencies and organization with recognized authority, expertise, or oversight in the areas addressed by these guidelines. 
Despite not being directly enforceable, as has been argued in prior work~\cite{haeder2020policies,haim2025binding,parrillo2019federal}, guidance documents can often act as \emph{soft laws} and be binding in \emph{practice}. 
Organizations often comply in anticipation of enforcement, adjusting behavior before rules are binding \citep{kim2026antidiversity}.
Agencies have long used guidance documents to implement major statutes~\cite{bagley2014essential,haeder2020policies}; in some cases guidance later became binding, as when the Department of Justice adopted accessibility standards of Web Content Accessibility Guidelines (WCAG) as the American Disabilities Act (ADA) Title II~\cite{doj2022guidance,doj2024finalrule}.
Because they are less procedural, and more flexible, governance by guidance can be a useful way for agencies to \emph{``respond more nimbly to changing political circumstances or technical and scientific innovations''} ~\cite{haeder2020policies}, such as developments around AI. 

In the past three years, more than half of U.S. states have responded to AI-related technological changes by publishing formal guidance documents for K-12 education. The U.S. federalism and decentralization of education governance provides affordances and opportunities for states to act as \emph{laboratories of democracy}~\cite{USStateLocalGovernments2007}: they can independently experiment with different approaches, while also allowing other jurisdictions to observe, learn, and possibly adopt. In this view, state AI guidance documents represent an uncoordinated yet parallel sensemaking process of largely independent state education departments responding to a shared technological disruption. These documents are not simply informational guidance but can be understood as \emph{governance artifacts}~\cite{strauss2019domesticating,rasheedpolicy} that encode state agencies' priorities, expectations, and illuminate several other dimensions of governance. 

In this study, we use a mixed-methods approach and present a systematic content analysis of AI guidance documents from the 35 U.S. states and Puerto Rico comprising a total of 1,539 pages. We compare the priorities set in these documents, and using eight other datasets, identify their relationship to the demographic and administrative capacity of states. We further trace 5{,}351 references in these documents to 328 distinct sources, including 61 citations to other states, to study their epistemic grounding, and possible patterns of policy diffusion, i.e., the spread and adoption of ideas and regulations across jurisdictions. Our study is guided by the following research questions:

\begin{itemize}[topsep=0pt, leftmargin=16pt]
    \item \textbf{{RQ1: How do the states compare in their governance priorities in terms of topics and audiences?}}
    \item \textbf{{RQ2: Are state characteristics of demographics (e.g., percentage of rural populations, english language learners), fiscal and infrastructural capacities (e.g., broadband access, state education funding), and political leanings associated with state priorities identified in the guidance documents?}}
    \item \textbf{{RQ3: What sources do states cite, and who writes their guidance?}}
\end{itemize}

\section{Background}
Under federalism, states are the primary governors of education in the United States, including how schools adopt new technologies. Yet, states have not always had the capacity to respond to rapid technological changes. For example, during COVID-19, states did not have enough capacity or governmental infrastructure to cope with fast-moving technological change like remote learning, and often relied on for-profit organizations for resources and tools~\cite{willse2024state}. Generative AI poses a similar challenge. A mid-2023 study surveying 102 high school principals and university provosts found that a majority of schools lacked guidelines for responsible integration of AI, with existing policies narrowly focused on plagiarism, and lacking issues of bias and transparency~\cite{ghimire2024guidelines}. The earliest reported published AI guidance for K-12 education was by the state of Oregon in August of 2023~\cite{pate2025aiguidance} followed by California in September of 2023~\cite{merod2023aiguidance}. Since then, 33 more states and Puerto Rico have followed suit.

A growing body of work is dedicated to the study of these AI policy and guidance documents as well as the context around it. A Delphi study of 24 international experts created a normative framework of 8 key priority areas for what a guidance document should cover, including academic integrity, equitable access, and responsible use~\cite{crompton2026framework}. Our study complements this work by examining what guidances' actually prioritize.
Two contemporaneous studies are the closest to ours: ~\citet{mccune2026fragile} coded state documents with the CAPE framework (Capacity, Access, Participation, Experience). ~\citet{rasheedpolicy} applied qualitative content analysis to state guidance documents. Both found that states frame AI primarily as a risk to mitigate rather than a tool to integrate into classrooms. Rasheed also found that ethics and privacy are covered consistently, while equity and capacity building receive little attention. Rasheed attributes the shared vocabulary across guidance documents to policy diffusion.
In our study, we code guidance documents across more granular categories within similar dimensions as Rasheed. We further test the relationship of guidance priorities with state demographics and capacities. We also investigate the authors of these documents and their affiliations.

Finally, policy diffusion literature~\cite{walker1969diffusion} has shown that states often emulate peers they regard as legitimate reference points, as well as rely on third-party organizations for standards~\cite{rice2023qualitative} and resources~\cite{willse2024state}. To trace such influence empirically, prior work has used citations as evidence of the transmission of ideas, science, and legal logic between state courts~\cite{hinkle2015transmission}, and policy documents~\cite{cabral2024mapping,ramirez2026government,yin2021coevolution}. 
We use citation analysis to understand if states learn from one another and/or share sources of epistemic grounding when it comes to AI guidance for K-12 education. 

\section{Methodology}

\subsection{Data Collection}

\subsubsection{AI Guidance Documents by U.S. State Education Departments} 
\label{sec:data-collection-guidances}
To collect state-wide AI guidance documents for K-12 education in the U.S., we use Ballotpedia,\footnote{\url{https://ballotpedia.org/ai_guidance_issued_by_state_departments_of_education}} a nonpartisan online political encyclopedia of American politics and elections. As of 31st August 2026, Ballotpedia contains a database of AI guidance links by 35 states. We also collected Puerto Rico's guidance from the AI for Education site.\footnote{\url{https://www.aiforeducation.io/}} For each guidance document, we verified the latest version from the official state websites. We limit our search to guidance documents issued by state departments of education. For example, Northern Arizona University has published guidance for K-12 usage,\footnote{\url{https://in.nau.edu/wp-content/uploads/sites/359/2025/05/AZ-GenAI-Guidance-v26.01.pdf}} but it is not part of our dataset. We also check if the state has issued a template or model policy (i.e., ready-made policy language that school districts can adopt or adapt) alongside guidance to form their own policies, and if so, we collect those as well. For states where we could not find guidance documents, we confirmed their unavailability from the NIAIE State AI Readiness Index Report~\cite{niaie2025stateAI} and by conducting manual search. In total, we identify and verify 36 state guidances (including Puerto Rico). Most documents are hosted as *.pdf or *.doc files. Some are hosted as interactive websites such as from Connecticut, Maine, New Jersey, North Dakota, and Ohio. In order to process these documents for analysis, we convert these webpages to PDFs for analysis. We will provide the processed corpus of these guidance documents, and metadata, as openly licensed data for reuse upon publication. 
 
\subsubsection{External Data Sources on State Demographics and Resources} To identify if there is any relationship between characteristics of states and their priorities in guidances (RQ2), we identify eight additional variables using three external sources of data. These include (i) U.S. Census Bureau (percentage of rural population, percentage of households having broadband subscription, and per pupil spending), (ii) National Center for Education Statistics (percentages of English language learners, population identifying as American Indian and Alaska Native (AI/AN), children served under Disabilities Education Act, and public school students eligible for free reduced-price lunch), and (iii) Ballotpedia (political leaning defined as the percentage of all seats across both legislative chambers combined that are held by Democrats). The data collected from these sources is the latest available and do not align exactly with the year of publication of the guidance documents. We do not expect this to affect our analysis as relative positions of states are unlikely to have shifted meaningfully within this period to affect our rank-based statistical test. For additional details related to exact sources, years, and tables, see Appendix Table~\ref{tab:rq2-data-sources}.  

\subsection{RQ1: Content Analysis}
\label{sec:content_analysis}
For RQ1, we utilize content analysis to identify which topics state guidance documents cover and whom they address. We focus on topic priorities covered by documents, rather than guidance attitudes towards AI, which has been explored in previous work~\cite{mccune2026fragile,rasheedpolicy}. We conduct a three-stage inductive and iterative content analysis process. First, two authors reviewed three documents using exploratory open coding and agreed upon a set of topics to track across documents, creating codes across seven categories: Document metadata, Document structure, Audience, Pedagogy \& Instruction, Actionability, Workforce \& career readiness, and Responsible use \& equity (see Appendix~\ref{sec:rq1_coding} for full set of questions). Second, the two authors independently coded all 36 documents with 86.3\% agreement (Cohen's $\kappa=0.73$ on 28 binary questions; substantial agreement~\cite{landis1977measurement}), taking informal notes throughout; disagreements were resolved through discussion and also led to updating or merging of codes, with overarching categories remaining the same. Third, Claude Sonnet 5 was used as an additional consistency check. We used a few-shot prompting strategy where we provided our codebook to an LLM. Two authors then manually reviewed every proposed coding change before accepting or rejecting it. Out of 145 of 1,080 cells flagged by Claude, adjudication upheld Claude's code decision in 41.4\% (n=60) of the highlighted cases. This step was added as our work relies on whether or not specific topics are discussed across states, and we did not want to inadvertently miss any important details or make claims that were untrue (see Appendix~\ref{sec:rq1_coding} for more details and data on Claude agreement with human coders). 

\subsection{RQ2: Hypothesis Testing}
\label{sec:hypothesis_testing}
For RQ2, we test the relationship of state priorities in guidance documents with state characteristics (demographics, state capacities, partisanship) using a set of formal hypotheses. We state and test a specific set of hypotheses related to how these state characteristics should relate to the priorities states express in the documents. Throughout, we test each state characteristic against each guidance dimension with a two-sided Mann-Whitney $U$ test, comparing the characteristic between documents that do and do not address the dimension, and report group medians and $p$-values. 

\subsection{RQ3: Citation and Contributor Analysis}
\label{sec:rq3-methods}

For RQ3, we examine the sources state guidance relies on, asking whose knowledge the guidance treats as reputable and trustworthy~\cite{Zagzebski}. We use citation analysis, treating citations as a marker of epistemic authority~\cite{beaver2004collaborative}. A guidance document confers authority in two ways: by citing outside works, such as laws, federal agencies, nonprofits, research centers, and other states' guidance, and by crediting the people and organizations who prepared the guidance itself. We refer to the former as \emph{epistemic sources} and to the latter as \emph{contributors}. To this end, we extract three kinds of information: First, \emph{cited sources} are outside organizations, laws, or named frameworks that a guidance document links to or names. We match each document's links and text against a dictionary of known sources, which identifies the source and its sector (e.g., a link to \texttt{teachai.org} or the name ``TeachAI'' is matched to TeachAI, a nonprofit, and ``FERPA'' to the Family Educational Rights and Privacy Act, a law). We further classify each source into one of ten types (see Appendix Table~\ref{tab:rq3-types}), eight of which follow the actors identified in prior policy transfer research~\cite{stone2004transfer, marsden2011policy}. Second, \emph{peer citations} are citations to other peer states' departments of education or to their guidances or resources. We classify peer citations into two types: An \emph{explicit adoption} occurs when the citing document takes something specific from the other state and acknowledges it, whether by reproducing or adapting its text, using one of its named models, roadmaps, or policy templates, or following one of its recommendations. Wisconsin, for instance, presents material as ``borrowed from'' Washington. On the other hand, a \emph{reference} refers to the document citing other state's work, as a reference-list entry, a footnote, or an item in a list of resources, without taking anything from it. Third, \emph{contributors} are the people, organizations, and group statements credited in each document. Each contributor was coded by their \emph{affiliation}, i.e., the organization printed alongside their name (state department of education, K-12 district or school, higher education, nonprofit, association, company or vendor, other government, or other), and the \emph{contributor group}, i.e., the role indicated by their title (teachers, administrators, higher education, policymakers, students, industry or vendors, parents or families, IT or technology staff, community, or librarians). These codes and a first draft of the source dictionary were produced with the assistance of Claude Code (Opus 4.8), and one of the authors reviewed all codes and dictionary mappings. Appendix Table~\ref{tab:rq3-entities} summarizes the three types of information we extract with examples, and Appendix~\ref{app:rq3-sources} further details the extraction, classification, and validation procedures.

\section{Findings}

\subsection{RQ1: How do states compare in their governance priorities in terms of topics and audiences?}
\definecolor{FamMeta}{HTML}{FCD5CE}
\definecolor{FamStruct}{HTML}{D8E2DC}
\definecolor{FamAud}{HTML}{F7E1D7}
\definecolor{FamPed}{HTML}{FFD7BA}
\definecolor{FamAct}{HTML}{BEE1E6}
\definecolor{FamWork}{HTML}{FFE5EC}
\definecolor{FamResp}{HTML}{CDDAFD}
\providecommand{\famsw}[1]{\raisebox{-0.15ex}{\setlength{\fboxsep}{0pt}\fcolorbox{black!35}{#1}{\rule{0pt}{1.3ex}\rule{1.3ex}{0pt}}}}
\begin{table*}[htbp]
\centering\scriptsize
\caption{Coverage of U.S.\ state K--12 AI guidance documents across coding dimensions, grouped into seven families (colors). \CIRCLE\ = addressed, \Circle\ = not addressed; the bottom \textbf{\% Yes} row gives the share of states that address each dimension. Shorthand labels are defined in Table~\ref{tab:coding-legend}.}
\label{tab:coding-grid}
\setlength{\tabcolsep}{3pt}\renewcommand{\arraystretch}{1.1}
\resizebox{\textwidth}{!}{%
\begin{tabular}{lr|>{\columncolor{FamMeta!35}}c>{\columncolor{FamMeta!35}}c>{\columncolor{FamMeta!35}}c>{\columncolor{FamMeta!35}}c|>{\columncolor{FamStruct!35}}c>{\columncolor{FamStruct!35}}c>{\columncolor{FamStruct!35}}c|>{\columncolor{FamAud!35}}c>{\columncolor{FamAud!35}}c>{\columncolor{FamAud!35}}c>{\columncolor{FamAud!35}}c>{\columncolor{FamAud!35}}c|>{\columncolor{FamPed!35}}c>{\columncolor{FamPed!35}}c>{\columncolor{FamPed!35}}c|>{\columncolor{FamAct!35}}c>{\columncolor{FamAct!35}}c>{\columncolor{FamAct!35}}c>{\columncolor{FamAct!35}}c|>{\columncolor{FamWork!35}}c|>{\columncolor{FamResp!35}}c>{\columncolor{FamResp!35}}c>{\columncolor{FamResp!35}}c>{\columncolor{FamResp!35}}c>{\columncolor{FamResp!35}}c>{\columncolor{FamResp!35}}c>{\columncolor{FamResp!35}}c>{\columncolor{FamResp!35}}c>{\columncolor{FamResp!35}}c>{\columncolor{FamResp!35}}c}
\toprule
State & \rotatebox[origin=l]{90}{\# pages} & \cellcolor{FamMeta}\rotatebox[origin=l]{90}{Version history} & \cellcolor{FamMeta}\rotatebox[origin=l]{90}{Year (recent)} & \cellcolor{FamMeta}\rotatebox[origin=l]{90}{Month (recent)} & \cellcolor{FamMeta}\rotatebox[origin=l]{90}{AI self-disclosure} & \cellcolor{FamStruct}\rotatebox[origin=l]{90}{Guiding principles} & \cellcolor{FamStruct}\rotatebox[origin=l]{90}{Glossary / definitions} & \cellcolor{FamStruct}\rotatebox[origin=l]{90}{Review cycle} & \cellcolor{FamAud}\rotatebox[origin=l]{90}{Administrators} & \cellcolor{FamAud}\rotatebox[origin=l]{90}{Teachers} & \cellcolor{FamAud}\rotatebox[origin=l]{90}{Students} & \cellcolor{FamAud}\rotatebox[origin=l]{90}{Families / parents} & \cellcolor{FamAud}\rotatebox[origin=l]{90}{IT / tech staff} & \cellcolor{FamPed}\rotatebox[origin=l]{90}{Professional development} & \cellcolor{FamPed}\rotatebox[origin=l]{90}{Assessment redesign} & \cellcolor{FamPed}\rotatebox[origin=l]{90}{Personalized learning} & \cellcolor{FamAct}\rotatebox[origin=l]{90}{Roadmap / timeline} & \cellcolor{FamAct}\rotatebox[origin=l]{90}{Procurement / implementation} & \cellcolor{FamAct}\rotatebox[origin=l]{90}{Funding / budget} & \cellcolor{FamAct}\rotatebox[origin=l]{90}{Vendor evaluation} & \cellcolor{FamWork}\rotatebox[origin=l]{90}{Workforce / jobs / upskilling} & \cellcolor{FamResp}\rotatebox[origin=l]{90}{Deepfakes / synthetic media} & \cellcolor{FamResp}\rotatebox[origin=l]{90}{Equity / digital divide} & \cellcolor{FamResp}\rotatebox[origin=l]{90}{Algorithmic bias} & \cellcolor{FamResp}\rotatebox[origin=l]{90}{English language learners} & \cellcolor{FamResp}\rotatebox[origin=l]{90}{Rural / underserved} & \cellcolor{FamResp}\rotatebox[origin=l]{90}{Tribal / Indigenous} & \cellcolor{FamResp}\rotatebox[origin=l]{90}{Accessibility \& disability} & \cellcolor{FamResp}\rotatebox[origin=l]{90}{Academic integrity} & \cellcolor{FamResp}\rotatebox[origin=l]{90}{Data privacy \& governance} & \cellcolor{FamResp}\rotatebox[origin=l]{90}{AI companion apps} \\
\midrule
Alabama & 13 & \CIRCLE & {\tiny 2024} & {\tiny Jun} & \Circle & \CIRCLE & \CIRCLE & \Circle & \CIRCLE & \CIRCLE & \CIRCLE & \Circle & \Circle & \CIRCLE & \Circle & \Circle & \Circle & \CIRCLE & \Circle & \CIRCLE & \CIRCLE & \Circle & \CIRCLE & \CIRCLE & \Circle & \Circle & \Circle & \Circle & \CIRCLE & \CIRCLE & \Circle \\
Alaska & 15 & \Circle & $\cdot$ & $\cdot$ & \CIRCLE & \CIRCLE & \CIRCLE & \Circle & \CIRCLE & \CIRCLE & \Circle & \CIRCLE & \Circle & \CIRCLE & \Circle & \CIRCLE & \Circle & \CIRCLE & \CIRCLE & \CIRCLE & \CIRCLE & \Circle & \CIRCLE & \CIRCLE & \Circle & \CIRCLE & \Circle & \Circle & \CIRCLE & \CIRCLE & \Circle \\
California & 57 & \Circle & {\tiny 2026} & {\tiny Jan} & \Circle & \Circle & \CIRCLE & \Circle & \CIRCLE & \CIRCLE & \Circle & \CIRCLE & \CIRCLE & \CIRCLE & \CIRCLE & \CIRCLE & \CIRCLE & \CIRCLE & \CIRCLE & \CIRCLE & \CIRCLE & \CIRCLE & \CIRCLE & \CIRCLE & \CIRCLE & \CIRCLE & \Circle & \CIRCLE & \CIRCLE & \CIRCLE & \CIRCLE \\
Colorado & 55 & \Circle & {\tiny 2025} & {\tiny Aug} & \Circle & \CIRCLE & \Circle & \CIRCLE & \CIRCLE & \CIRCLE & \Circle & \CIRCLE & \Circle & \CIRCLE & \CIRCLE & \CIRCLE & \CIRCLE & \CIRCLE & \CIRCLE & \CIRCLE & \CIRCLE & \CIRCLE & \CIRCLE & \CIRCLE & \CIRCLE & \CIRCLE & \Circle & \CIRCLE & \CIRCLE & \CIRCLE & \Circle \\
Connecticut & 13 & \Circle & $\cdot$ & $\cdot$ & \Circle & \Circle & \Circle & \Circle & \CIRCLE & \CIRCLE & \Circle & \CIRCLE & \Circle & \CIRCLE & \Circle & \CIRCLE & \Circle & \CIRCLE & \Circle & \CIRCLE & \Circle & \CIRCLE & \CIRCLE & \CIRCLE & \Circle & \Circle & \Circle & \Circle & \CIRCLE & \CIRCLE & \CIRCLE \\
Delaware & 28 & \CIRCLE & {\tiny 2024} & {\tiny Jun} & \Circle & \CIRCLE & \CIRCLE & \CIRCLE & \CIRCLE & \CIRCLE & \Circle & \CIRCLE & \Circle & \CIRCLE & \CIRCLE & \CIRCLE & \CIRCLE & \CIRCLE & \Circle & \CIRCLE & \CIRCLE & \CIRCLE & \CIRCLE & \CIRCLE & \Circle & \Circle & \Circle & \CIRCLE & \CIRCLE & \CIRCLE & \Circle \\
Georgia & 24 & \Circle & {\tiny 2025} & {\tiny Jun} & \Circle & \CIRCLE & \CIRCLE & \CIRCLE & \CIRCLE & \CIRCLE & \Circle & \CIRCLE & \CIRCLE & \CIRCLE & \Circle & \CIRCLE & \CIRCLE & \CIRCLE & \CIRCLE & \CIRCLE & \Circle & \Circle & \CIRCLE & \CIRCLE & \Circle & \Circle & \Circle & \CIRCLE & \CIRCLE & \CIRCLE & \Circle \\
Hawaii & 2 & \Circle & {\tiny 2024} & {\tiny May} & \Circle & \Circle & \Circle & \CIRCLE & \CIRCLE & \CIRCLE & \Circle & \Circle & \Circle & \CIRCLE & \CIRCLE & \CIRCLE & \Circle & \CIRCLE & \Circle & \Circle & \Circle & \Circle & \CIRCLE & \CIRCLE & \Circle & \Circle & \Circle & \Circle & \CIRCLE & \CIRCLE & \Circle \\
Illinois & 408 & \Circle & {\tiny 2026} & {\tiny Jun} & \CIRCLE & \CIRCLE & \CIRCLE & \Circle & \CIRCLE & \CIRCLE & \CIRCLE & \CIRCLE & \Circle & \CIRCLE & \CIRCLE & \CIRCLE & \CIRCLE & \CIRCLE & \CIRCLE & \CIRCLE & \CIRCLE & \CIRCLE & \CIRCLE & \CIRCLE & \CIRCLE & \CIRCLE & \Circle & \CIRCLE & \CIRCLE & \CIRCLE & \CIRCLE \\
Indiana & 5 & \Circle & $\cdot$ & $\cdot$ & \Circle & \Circle & \Circle & \Circle & \CIRCLE & \CIRCLE & \Circle & \CIRCLE & \Circle & \Circle & \Circle & \CIRCLE & \Circle & \Circle & \Circle & \Circle & \CIRCLE & \Circle & \CIRCLE & \CIRCLE & \Circle & \Circle & \Circle & \Circle & \CIRCLE & \CIRCLE & \Circle \\
Kentucky & 13 & \Circle & {\tiny 2025} & {\tiny Dec} & \CIRCLE & \CIRCLE & \CIRCLE & \Circle & \CIRCLE & \CIRCLE & \Circle & \Circle & \Circle & \CIRCLE & \Circle & \CIRCLE & \Circle & \Circle & \Circle & \CIRCLE & \CIRCLE & \Circle & \Circle & \CIRCLE & \Circle & \Circle & \Circle & \Circle & \CIRCLE & \CIRCLE & \Circle \\
Louisiana & 18 & \Circle & {\tiny 2024} & {\tiny Aug} & \Circle & \CIRCLE & \CIRCLE & \Circle & \CIRCLE & \CIRCLE & \Circle & \CIRCLE & \CIRCLE & \CIRCLE & \CIRCLE & \CIRCLE & \Circle & \CIRCLE & \Circle & \CIRCLE & \CIRCLE & \Circle & \Circle & \CIRCLE & \CIRCLE & \Circle & \Circle & \CIRCLE & \CIRCLE & \CIRCLE & \Circle \\
Maine & 50 & \Circle & {\tiny 2025} & $\cdot$ & \Circle & \CIRCLE & \CIRCLE & \Circle & \CIRCLE & \CIRCLE & \Circle & \Circle & \Circle & \CIRCLE & \CIRCLE & \CIRCLE & \CIRCLE & \Circle & \CIRCLE & \CIRCLE & \CIRCLE & \Circle & \CIRCLE & \CIRCLE & \Circle & \Circle & \Circle & \CIRCLE & \CIRCLE & \CIRCLE & \Circle \\
Maryland & 16 & \Circle & {\tiny 2026} & {\tiny Jun} & \Circle & \Circle & \CIRCLE & \CIRCLE & \CIRCLE & \CIRCLE & \CIRCLE & \CIRCLE & \Circle & \CIRCLE & \CIRCLE & \CIRCLE & \Circle & \CIRCLE & \Circle & \CIRCLE & \CIRCLE & \Circle & \CIRCLE & \CIRCLE & \CIRCLE & \CIRCLE & \Circle & \CIRCLE & \CIRCLE & \CIRCLE & \Circle \\
Massachusetts & 74 & \Circle & $\cdot$ & $\cdot$ & \Circle & \CIRCLE & \CIRCLE & \Circle & \CIRCLE & \CIRCLE & \Circle & \CIRCLE & \CIRCLE & \CIRCLE & \CIRCLE & \CIRCLE & \CIRCLE & \CIRCLE & \CIRCLE & \CIRCLE & \CIRCLE & \CIRCLE & \CIRCLE & \CIRCLE & \CIRCLE & \CIRCLE & \Circle & \CIRCLE & \CIRCLE & \CIRCLE & \Circle \\
Michigan & 11 & \Circle & $\cdot$ & $\cdot$ & \CIRCLE & \CIRCLE & \Circle & \Circle & \CIRCLE & \CIRCLE & \Circle & \Circle & \CIRCLE & \CIRCLE & \CIRCLE & \Circle & \CIRCLE & \CIRCLE & \CIRCLE & \CIRCLE & \Circle & \Circle & \CIRCLE & \CIRCLE & \Circle & \Circle & \Circle & \CIRCLE & \CIRCLE & \CIRCLE & \Circle \\
Minnesota & 6 & \Circle & {\tiny 2024} & {\tiny Jul} & \CIRCLE & \CIRCLE & \Circle & \Circle & \CIRCLE & \CIRCLE & \Circle & \CIRCLE & \Circle & \CIRCLE & \CIRCLE & \CIRCLE & \Circle & \Circle & \Circle & \Circle & \CIRCLE & \Circle & \CIRCLE & \CIRCLE & \Circle & \Circle & \Circle & \CIRCLE & \CIRCLE & \CIRCLE & \Circle \\
Mississippi & 26 & \Circle & $\cdot$ & $\cdot$ & \Circle & \Circle & \CIRCLE & \Circle & \CIRCLE & \CIRCLE & \CIRCLE & \CIRCLE & \Circle & \CIRCLE & \CIRCLE & \CIRCLE & \CIRCLE & \CIRCLE & \Circle & \Circle & \CIRCLE & \Circle & \CIRCLE & \CIRCLE & \Circle & \Circle & \Circle & \CIRCLE & \CIRCLE & \CIRCLE & \Circle \\
Missouri & 22 & \CIRCLE & {\tiny 2025} & $\cdot$ & \Circle & \CIRCLE & \CIRCLE & \Circle & \CIRCLE & \CIRCLE & \Circle & \CIRCLE & \CIRCLE & \CIRCLE & \CIRCLE & \CIRCLE & \Circle & \CIRCLE & \CIRCLE & \CIRCLE & \CIRCLE & \CIRCLE & \CIRCLE & \CIRCLE & \Circle & \Circle & \Circle & \CIRCLE & \CIRCLE & \CIRCLE & \Circle \\
Montana & 22 & \CIRCLE & {\tiny 2026} & {\tiny Aug} & \CIRCLE & \CIRCLE & \CIRCLE & \CIRCLE & \CIRCLE & \CIRCLE & \Circle & \CIRCLE & \Circle & \CIRCLE & \Circle & \Circle & \CIRCLE & \CIRCLE & \CIRCLE & \CIRCLE & \CIRCLE & \CIRCLE & \CIRCLE & \CIRCLE & \CIRCLE & \CIRCLE & \CIRCLE & \CIRCLE & \CIRCLE & \CIRCLE & \CIRCLE \\
Nevada & 54 & \Circle & {\tiny 2025} & {\tiny Mar} & \Circle & \CIRCLE & \Circle & \CIRCLE & \CIRCLE & \CIRCLE & \CIRCLE & \CIRCLE & \Circle & \CIRCLE & \CIRCLE & \CIRCLE & \Circle & \Circle & \CIRCLE & \CIRCLE & \CIRCLE & \Circle & \CIRCLE & \CIRCLE & \CIRCLE & \CIRCLE & \Circle & \CIRCLE & \CIRCLE & \CIRCLE & \Circle \\
New Jersey & 3 & \Circle & $\cdot$ & $\cdot$ & \Circle & \CIRCLE & \CIRCLE & \Circle & \Circle & \Circle & \Circle & \Circle & \Circle & \Circle & \Circle & \Circle & \Circle & \Circle & \Circle & \Circle & \CIRCLE & \Circle & \Circle & \Circle & \Circle & \Circle & \Circle & \Circle & \Circle & \Circle & \Circle \\
New Mexico & 31 & \CIRCLE & {\tiny 2025} & {\tiny May} & \CIRCLE & \CIRCLE & \CIRCLE & \CIRCLE & \CIRCLE & \CIRCLE & \Circle & \CIRCLE & \Circle & \CIRCLE & \CIRCLE & \CIRCLE & \Circle & \CIRCLE & \Circle & \CIRCLE & \CIRCLE & \CIRCLE & \CIRCLE & \CIRCLE & \Circle & \CIRCLE & \CIRCLE & \CIRCLE & \CIRCLE & \CIRCLE & \Circle \\
North Carolina & 106 & \CIRCLE & {\tiny 2026} & {\tiny Aug} & \CIRCLE & \CIRCLE & \CIRCLE & \CIRCLE & \CIRCLE & \CIRCLE & \CIRCLE & \CIRCLE & \CIRCLE & \CIRCLE & \CIRCLE & \CIRCLE & \CIRCLE & \CIRCLE & \Circle & \CIRCLE & \CIRCLE & \CIRCLE & \CIRCLE & \CIRCLE & \CIRCLE & \CIRCLE & \Circle & \CIRCLE & \CIRCLE & \CIRCLE & \CIRCLE \\
North Dakota & 19 & \Circle & $\cdot$ & $\cdot$ & \Circle & \Circle & \CIRCLE & \Circle & \CIRCLE & \CIRCLE & \Circle & \CIRCLE & \CIRCLE & \CIRCLE & \CIRCLE & \CIRCLE & \CIRCLE & \CIRCLE & \Circle & \CIRCLE & \CIRCLE & \Circle & \CIRCLE & \CIRCLE & \CIRCLE & \CIRCLE & \Circle & \CIRCLE & \CIRCLE & \CIRCLE & \Circle \\
Ohio & 105 & \Circle & {\tiny 2026} & {\tiny Jan} & \Circle & \CIRCLE & \CIRCLE & \Circle & \CIRCLE & \CIRCLE & \Circle & \CIRCLE & \Circle & \CIRCLE & \CIRCLE & \CIRCLE & \CIRCLE & \CIRCLE & \CIRCLE & \CIRCLE & \CIRCLE & \CIRCLE & \Circle & \CIRCLE & \CIRCLE & \Circle & \Circle & \CIRCLE & \CIRCLE & \CIRCLE & \Circle \\
Oklahoma & 30 & \CIRCLE & {\tiny 2026} & {\tiny Jul} & \Circle & \CIRCLE & \CIRCLE & \CIRCLE & \CIRCLE & \CIRCLE & \Circle & \CIRCLE & \CIRCLE & \CIRCLE & \CIRCLE & \CIRCLE & \CIRCLE & \CIRCLE & \CIRCLE & \CIRCLE & \CIRCLE & \CIRCLE & \CIRCLE & \CIRCLE & \CIRCLE & \CIRCLE & \CIRCLE & \CIRCLE & \CIRCLE & \CIRCLE & \CIRCLE \\
Oregon & 42 & \Circle & {\tiny 2026} & {\tiny Jun} & \Circle & \CIRCLE & \Circle & \Circle & \CIRCLE & \CIRCLE & \Circle & \CIRCLE & \CIRCLE & \CIRCLE & \CIRCLE & \CIRCLE & \Circle & \CIRCLE & \Circle & \CIRCLE & \CIRCLE & \CIRCLE & \CIRCLE & \CIRCLE & \CIRCLE & \CIRCLE & \CIRCLE & \CIRCLE & \CIRCLE & \CIRCLE & \CIRCLE \\
Puerto Rico & 50 & \Circle & {\tiny 2024} & {\tiny Aug} & \Circle & \CIRCLE & \CIRCLE & \Circle & \CIRCLE & \CIRCLE & \Circle & \CIRCLE & \Circle & \CIRCLE & \CIRCLE & \CIRCLE & \CIRCLE & \Circle & \CIRCLE & \CIRCLE & \CIRCLE & \Circle & \CIRCLE & \CIRCLE & \Circle & \Circle & \Circle & \CIRCLE & \CIRCLE & \CIRCLE & \Circle \\
Rhode Island & 53 & \Circle & {\tiny 2025} & {\tiny Aug} & \CIRCLE & \CIRCLE & \CIRCLE & \CIRCLE & \CIRCLE & \CIRCLE & \Circle & \CIRCLE & \CIRCLE & \CIRCLE & \CIRCLE & \CIRCLE & \CIRCLE & \CIRCLE & \CIRCLE & \CIRCLE & \CIRCLE & \CIRCLE & \CIRCLE & \CIRCLE & \CIRCLE & \CIRCLE & \Circle & \CIRCLE & \CIRCLE & \CIRCLE & \Circle \\
Utah & 16 & \Circle & {\tiny 2026} & {\tiny Mar} & \Circle & \CIRCLE & \CIRCLE & \CIRCLE & \CIRCLE & \CIRCLE & \CIRCLE & \CIRCLE & \Circle & \Circle & \CIRCLE & \CIRCLE & \Circle & \Circle & \Circle & \CIRCLE & \CIRCLE & \CIRCLE & \CIRCLE & \CIRCLE & \CIRCLE & \Circle & \Circle & \CIRCLE & \CIRCLE & \CIRCLE & \Circle \\
Virginia & 5 & \Circle & $\cdot$ & $\cdot$ & \Circle & \CIRCLE & \Circle & \CIRCLE & \CIRCLE & \CIRCLE & \Circle & \CIRCLE & \Circle & \CIRCLE & \CIRCLE & \CIRCLE & \CIRCLE & \CIRCLE & \Circle & \Circle & \CIRCLE & \Circle & \Circle & \CIRCLE & \Circle & \Circle & \Circle & \Circle & \CIRCLE & \CIRCLE & \Circle \\
Washington & 54 & \CIRCLE & {\tiny 2024} & {\tiny Jul} & \CIRCLE & \CIRCLE & \CIRCLE & \CIRCLE & \CIRCLE & \CIRCLE & \CIRCLE & \CIRCLE & \CIRCLE & \CIRCLE & \CIRCLE & \CIRCLE & \CIRCLE & \CIRCLE & \CIRCLE & \CIRCLE & \CIRCLE & \CIRCLE & \CIRCLE & \CIRCLE & \CIRCLE & \CIRCLE & \Circle & \CIRCLE & \CIRCLE & \CIRCLE & \Circle \\
West Virginia & 44 & \CIRCLE & {\tiny 2025} & {\tiny Mar} & \CIRCLE & \CIRCLE & \CIRCLE & \CIRCLE & \CIRCLE & \CIRCLE & \Circle & \CIRCLE & \Circle & \CIRCLE & \CIRCLE & \CIRCLE & \CIRCLE & \CIRCLE & \Circle & \CIRCLE & \CIRCLE & \CIRCLE & \CIRCLE & \CIRCLE & \CIRCLE & \Circle & \Circle & \CIRCLE & \CIRCLE & \CIRCLE & \CIRCLE \\
Wisconsin & 40 & \CIRCLE & {\tiny 2024} & {\tiny Dec} & \Circle & \CIRCLE & \CIRCLE & \CIRCLE & \CIRCLE & \CIRCLE & \Circle & \CIRCLE & \Circle & \CIRCLE & \CIRCLE & \CIRCLE & \CIRCLE & \CIRCLE & \CIRCLE & \CIRCLE & \CIRCLE & \Circle & \CIRCLE & \CIRCLE & \Circle & \Circle & \CIRCLE & \CIRCLE & \CIRCLE & \CIRCLE & \Circle \\
Wyoming & 9 & \Circle & $\cdot$ & $\cdot$ & \Circle & \CIRCLE & \CIRCLE & \CIRCLE & \CIRCLE & \Circle & \Circle & \CIRCLE & \Circle & \CIRCLE & \Circle & \Circle & \Circle & \CIRCLE & \Circle & \CIRCLE & \Circle & \CIRCLE & \CIRCLE & \CIRCLE & \CIRCLE & \Circle & \Circle & \Circle & \CIRCLE & \CIRCLE & \Circle \\
\midrule
\textbf{\% Yes} & {\scriptsize --} & {\scriptsize\textbf{28}} & {\scriptsize --} & {\scriptsize --} & {\scriptsize\textbf{31}} & {\scriptsize\textbf{81}} & {\scriptsize\textbf{75}} & {\scriptsize\textbf{47}} & {\scriptsize\textbf{97}} & {\scriptsize\textbf{94}} & {\scriptsize\textbf{22}} & {\scriptsize\textbf{83}} & {\scriptsize\textbf{33}} & {\scriptsize\textbf{92}} & {\scriptsize\textbf{75}} & {\scriptsize\textbf{86}} & {\scriptsize\textbf{56}} & {\scriptsize\textbf{78}} & {\scriptsize\textbf{47}} & {\scriptsize\textbf{83}} & {\scriptsize\textbf{86}} & {\scriptsize\textbf{50}} & {\scriptsize\textbf{86}} & {\scriptsize\textbf{97}} & {\scriptsize\textbf{50}} & {\scriptsize\textbf{42}} & {\scriptsize\textbf{14}} & {\scriptsize\textbf{75}} & {\scriptsize\textbf{97}} & {\scriptsize\textbf{97}} & {\scriptsize\textbf{22}} \\
\bottomrule
\end{tabular}}
\par\smallskip{\footnotesize \famsw{FamMeta}~Document metadata \quad \famsw{FamStruct}~Document structure \quad \famsw{FamAud}~Audience \quad \famsw{FamPed}~Pedagogy \& instruction \quad \famsw{FamAct}~Actionability \quad \famsw{FamWork}~Workforce \& career readiness \quad \famsw{FamResp}~Responsible use \& equity}
\end{table*}

Table~\ref{tab:coding-grid} summarizes our coding across seven broad categories: Document Metadata, Document Structure, Audience, Pedagogy \& Instruction, Actionability, Workforce \& Career Readiness, and Responsible Use \& Equity (see Appendix Table~\ref{tab:coding-legend} for definitions). The sections below cover main findings in each category; we interpret results further in the discussion section.

\subsubsection{Document metadata and structure}
The categories of document metadata and document structure review the characteristics of the documents themselves, rather than the guidance they give. This is an important starting point to understand when guidance was released, what are the versions of the guidance, and if they used AI themselves. State guidance documents vary widely in length from 2 to 408 pages; a majority do not include versioning history (72\%) and less than half of the states commit to a review cycle. We discuss the importance of regular updates and clear versioning history in our Recommendations section in 5.5. Of the 36 jurisdictions, 11 mention some level of AI usage disclosure in the editing of the guidance documents; 75\% of the states include definitions or glossary sections, and 81\% include guiding principles. 

\subsubsection{Audience}
Coding under ``Audience'' reviews who the guidance is explicitly written for, showing who the states expect to be reading and acting upon the guidance. We find there is a large gap in who the guidance is given to. Throughout the documents, school administration (97\%) and teachers (94\%) are the main audience, with corresponding high coverage on instruction and pedagogy. However, students are addressed directly only in 22\% of the documents: Alabama, Illinois, Maryland, Mississippi, Nevada, North Carolina, Utah, and Washington. Similarly, though IT and tech staff will likely carry out data privacy protection and vendor evaluations, they are only considered as a direct audience of this guidance in 33\% of the documents.

\subsubsection{Pedagogy and Instruction}
Professional development is mentioned across most states (92\%). However, guidance does not elaborate on the format of professional development, how it fits into existing teacher or administration schedules, and what topics it should cover. Additionally, the needs of teachers are different. Thus, specific guidance on who should conduct the professional development, who should attend, and what it should achieve is missing. We observe similar issues with the idea of ``personalized learning,'' which is mentioned across 86\% of the documents as a use-case for AI and used to argue for the introduction of AI into the classroom. Yet, it is underspecified, with no examples of software or platforms that conduct personalized learning in practice. Some examples of this can be seen in:
\begin{itemize}
    \item \textit{"AI should be leveraged to enhance curriculum development, adaptive learning, and personalized education, catering to individual student needs." }(Kentucky, p9)
    \item \textit{"General Benefits of AI in Education: Personalized Learning: Adaptive learning platforms can tailor content, pacing, and scaffolding difficulty to individual student needs under human oversight."} (Oklahoma, p12)
   \item \textit{"AI can enhance educator productivity, streamline administrative tasks, and personalize learning."} (Maine, p29)
    \item \textit{"AI should be used to support engagement with complex ideas, provide appropriate scaffolding and feedback, and personalize learning in ways that respect student identity and experience."} (Maryland, p10)

\end{itemize}

\subsubsection{Actionability}
Actionability aims to measure if states provide roadmaps and timelines, as well as practical advice on procurement, implementation, funding, and vendor evaluation. This is important as, if states do not provide actions that stakeholders can take, guidance may be too abstract to enact change. One form of actionability is seen in Ohio's template policy for schools to create their own policies on AI usage, allowing schools to directly take this template and enact change. Another example of actionable advice related to funding and budgets is Rhode Island's statement to set a contractual not-to-exceed financial amount when paying for AI (Rhode Island, p47). Montana, Michigan and North Carolina's guidance includes rubrics for vendor evaluation and procurement that stakeholders can directly implement into their workflow. However, many states do not include practical advice on important tasks such as vendor evaluation, procurement, and post-implementation evaluations.  

\subsubsection{Workforce and Career readiness}
Workforce and career readiness aims to capture how documents use a future-driven approach to either justify or explain AI. Upskilling students for workforce and career readiness is mentioned in 86\% of guidance documents, and AI integration is often framed as an \emph{urgent} need to prepare students for an \emph{``AI-driven world''}.  Examples of such phrases include: 
\begin{itemize}
   \item \emph{``This guidance ensures Oklahoma students are prepared for an \textbf{AI-driven future}.''} (Oklahoma, p3)
    \item \emph{``Most higher paying \textbf{jobs will require AI literacy} and computer programming. Schools who do not teach these skills are not adequately preparing students for college and career readiness.''} (Rhode Island, p32)
    \item \emph{``[..] we must \textbf{urgently equip students} and workers with the skills, knowledge, and competencies to harness AI responsibly and effectively.} (Virginia, p1)
    \item \emph{``In today's digital world, \textbf{educators must embrace AI} in the classroom to prepare students for a technology-driven future.''} (Maine, p20)
   \item \emph{``Public education stands at the threshold of a new era defined by artificial intelligence (AI), demanding \textbf{urgent action} from educators and policymakers to prepare students for a future \textbf{transformed by rapidly evolving AI technologies}.''} (North Carolina, p6)
\end{itemize}

\subsubsection{Responsible use and equity}
Finally, we review topics related to responsible use and equity. Given the existing and emergent issues with responsible technology use related (such as bias, deepfakes, and equity of access) it's important to review which issues are (and are not) considered. We find some issues are consistently addressed, with bias, academic integrity, and data privacy covered in almost every document. Data privacy is mentioned in 97\% of documents, potentially reflecting the existence of ready-made student-privacy laws and compliance frameworks such as FERPA, COPPA, and so on, which are easily cited in the context of AI. 
Accessibility has a strong legal mandate (e.g., IDEA, Section 504, and ADA), and is one of the areas GenAI may help educators in transforming material to reach a broader swath of their student body; we find accessibility mentions in 75\% of the documents. Emergent issues that are borne from new affordances of generative AI (namely, deepfakes and AI companions) are less discussed and mentioned (50\% and 22\% respectively), mostly in documents released or updated in 2026. While equity of access to AI technology is mentioned as a consideration across most guidance documents (86\%), specific discussions of the marginalized populations i.e. rural/underserved, or tribal/indigenous communities that may need such access are not addressed widely (42\% and 14\% respectively).

\subsection{RQ2: Are state characteristics of demographics, fiscal and infrastructural capacities, and political leanings associated with state priorities identified in the guidance documents?}
We expect that characteristics of states, such as student demographics, fiscal capacity, and political lean, may associate with the prevalence or lack of specific priorities across guidances. We elaborate on these findings below. 

\subsubsection{\textbf{RQ2A (Demographic Responsiveness): Are demographic characteristics associated with state priorities in terms of demographic groups addressed in the guidance?}}
We test four hypothesis: States with larger shares of (i) rural students, (ii) English learners, (iii) Native students, and (iv) students served under disabilities act (IDEA) are more likely to address, respectively, rural or underserved learners, English learners, tribal or indigenous populations, and accessibility and disability.

We find no evidence of demographic responsiveness for three of the four groups. For each group, the share of that group in the state did not differ significantly between states that do and do not mention it (medians, reported as mention vs.\ no mention): rural or underserved learners (median 16.6\% vs. 28.1\%, $p=.18$), English-learners (median 9.4\% vs. 7.3\%, $p=.35$) and students with disabilities (median 15.2\% vs. 16.6\%, $p=.83$). The one exception is Native AI/AN representation. We did find that states that address tribal or indigenous population have higher Native enrollment than those that do not (median 10.3\% vs. 0.4\%, $p=.001$).

\subsubsection{\textbf{RQ2B (Capacity Building): Are state fiscal and infrastructure resources associated with state priorities for capacity building?}}
Following McDonnell and Elmore~\cite{mcdonnell1987getting}, we define capacity building as policy that invests in the material, intellectual, and human resources educators need to implement reforms over the long term, such as professional development and technical infrastructure
We test the following set of hypotheses: (a) states with higher broadband access and higher free or reduced price lunch share are each associated with higher chance of addressing equity or digital divide dimension, and (b) states with higher per-pupil spending are more likely to address, professional development, funding/budget, and upskilling. We do not find significant differences across groups that do and do not mention it (medians, reported as mention vs.\ no mention): Broadband access (median 91.8\% vs. 91.2\%, $p=.52$) and free or reduced-price lunch share (median 48.6\% vs. 54.6\%, $p=.87$) were both null. Likewise, per-pupil spending did not differ between states that do or do not provide guidance related to professional development guidance (median 17,181 vs. 13,622, $p=.48$), or funding and budget (median 17,785 vs. 17,102, $p=.61$). For upskilling, we find a near-threshold trend in the opposite direction: states that address it spend slightly less per pupil (median 17,100 vs. 20,521, $p=.09$). Overall, we find state capacity to be largely decoupled from the priorities of states within the guidance documents. 

\subsubsection{\textbf{RQ2C (Partisan Priorities): Is a state's political leaning associated with responsible use and equity-related priorities?}}

Here we test whether states with more Democratic legislatures are more likely to address responsible use and equity dimensions as described in Table~\ref{tab:coding-grid}. Three responsible-use dimensions---algorithmic bias, academic integrity, and data privacy and governance---are addressed by 35 out of 36 documents and therefore reflect bipartisan consensus with little variation to be tested. Among the the others that vary, we find no significant difference (all $p>.05$) across groups that do and do not mention it (medians, reported as mention vs.\ no mention): equity of access (median democratic share 49.1 vs. 32.6, $p=.45$), English language learners (median 49.8 vs. 48.0, $p=.93$), rural or underserved learners (median 62.5 vs. 38.1, $p=.051$), tribal or indigenous populations (median 45.5 vs. 49.1, $p=.70$), accessibility and disability (49.1 vs. 38.3, $p=.56$), deepfakes or synthetic media (median 59.4 vs. 45.5, $p=.48$), and AI companionship (median 49.8 vs. 48.0, $p=.95$). Overall, there seems to be no significant pattern of partisan division associated with guidance on equity or responsible AI use.

\subsection{RQ3: What sources do states cite, and who writes their guidance?}
\label{sec:rq3-results}

\begin{figure*}[t]
  \centering
  \includegraphics[width=\textwidth,keepaspectratio]{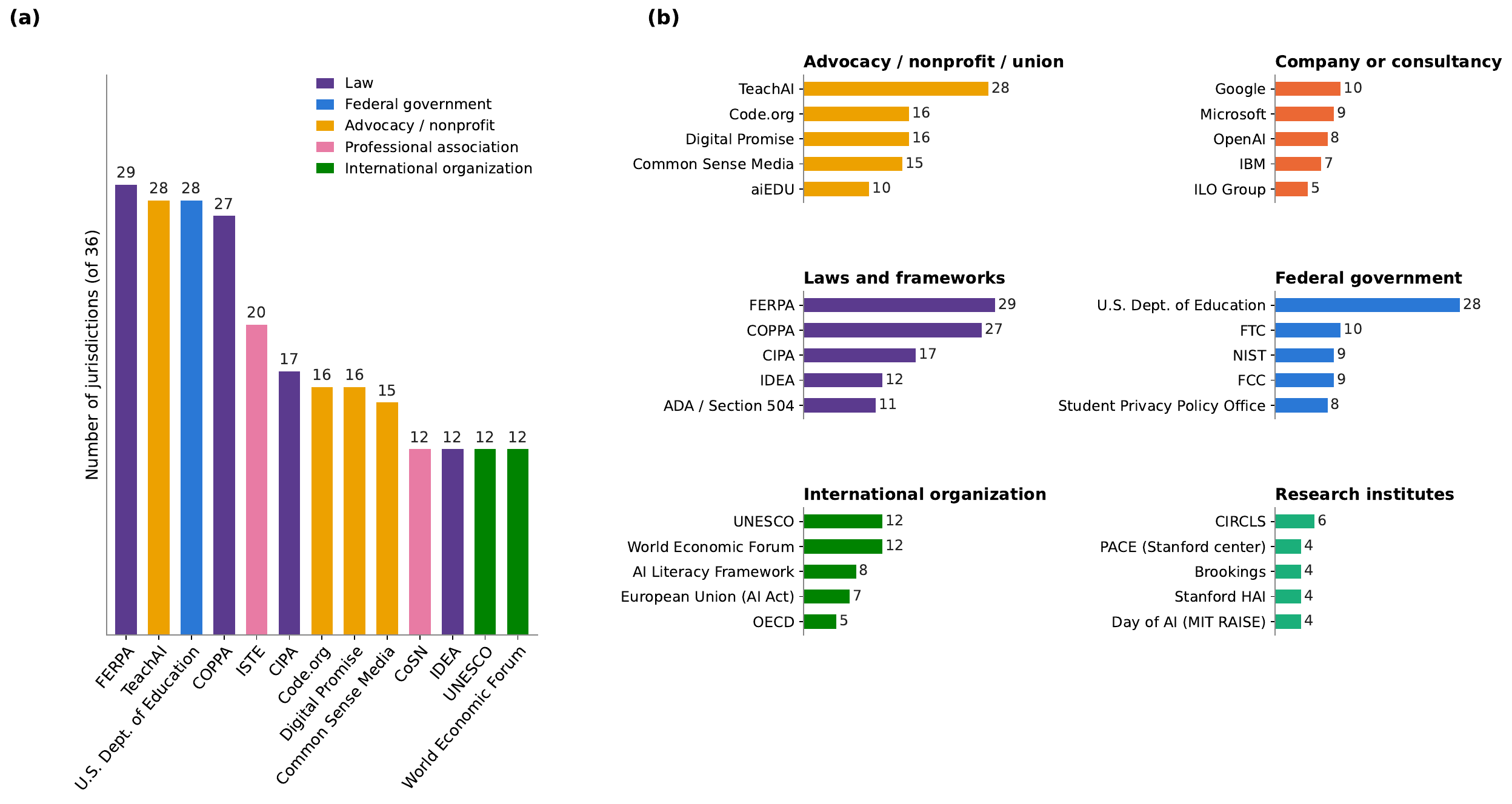}
  \caption{Most widely cited sources, colored by type of source. \textbf{(a)}~Sources cited by twelve or more of the 36 jurisdictions (35 states and Puerto Rico). \textbf{(b)}~The five most-cited sources within each sector.}
  \label{fig:rq3-sources}
\end{figure*}

If states act as laboratories that experiment and learn from one another, we expect their guidance should draw on varied sources, cite many different states, and be written by a broad range of people. If states instead rely on the same outside bodies, a few sources should appear in most documents, and citations to other states should concentrate on just a few documents. Here we identify common cited sources across state documents (RQ3A), patterns of states citing other peer states (RQ3B), and identify the contributors credited with preparing the guidance documents (RQ3C).

\subsubsection{\textbf{RQ3A (Cited Sources): What sources do states cite in their guidance?}}

Across the 36 documents, states cite 328 distinct sources, a median of 19 per state (Appendix Table~\ref{tab:rq3-types} breaks these down by type). Five sources are cited by at least half of the states: FERPA (29 states), TeachAI (28), the U.S.\ Department of Education (28), COPPA (27), and ISTE (20) (Figure~\ref{fig:rq3-sources}). Laws are the most consistently shared part among states: student privacy and disability laws such as FERPA, COPPA, CIPA, and IDEA are each cited by 12 to 29 states, which mirrors RQ1 findings, in which data privacy is addressed by 97\% of documents. Among non-government sources, the most widely cited are TeachAI, Code.org (16), and Digital Promise (16), followed by Common Sense Media (15). Beyond these, states cite largely different sources. Two-thirds of all sources (219 of 328) are cited by only one state, and no AI-specific framework is cited by more than eight (Appendix Figures~\ref{fig:laws} and~\ref{fig:frameworks}). On average, a pair of states shares about one in eight of the sources that either cites (mean Jaccard similarity 0.13), and states in the same census region are no more similar than states in different regions (difference $=-0.005$; permutation test, $p=.75$). Overall, the sources that states share consist largely of federal law and a cluster of connected nonprofits, with the remainder of sources being used uniquely rather than across states. Similarity across states thus runs through a common legal foundation and a few national organizations rather than through geography.

\subsubsection{\textbf{RQ3B (Peer Citation): Do states cite each other's guidance, and which states do they cite?}}

\begin{figure*}[t]
  \centering
  \includegraphics[width=\textwidth,keepaspectratio]{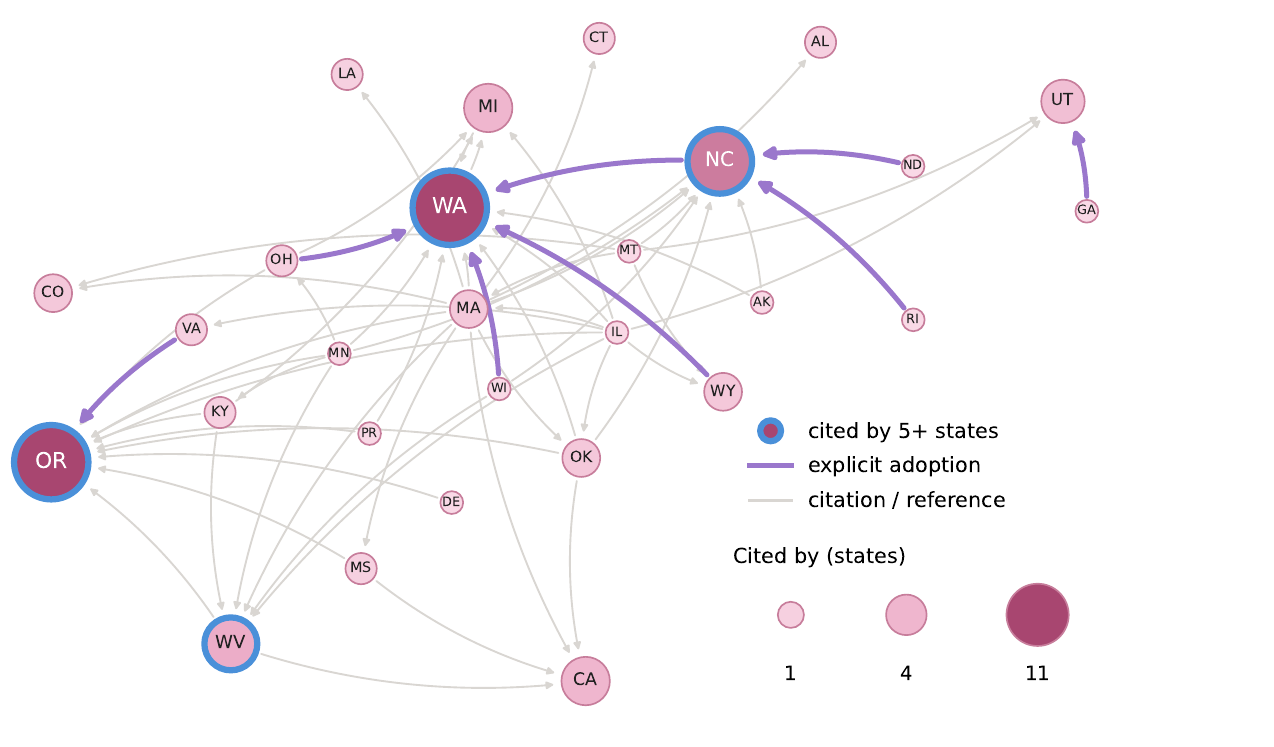}
  \caption{States citing other states' guidance (61 citations). Arrows indicate citation direction; node size and color indicate numbers of citations, and blue rings mark states cited by five or more. Purple arrows are the 8 citations coded as explicit adoptions of the other state's work. Fifteen states cite no other state (AL, CA, CO, CT, HI, IN, LA, MD, ME, MO, NJ, NM, NV, OR, UT) and appear only if another state cites them.}
  \label{fig:rq3-network}
\end{figure*}

Peer citation is common but concentrated to a few early states and mostly runs in one direction. 21 of the 36 states cite at least one other state, with a total of 61 peer-citations across documents (Figure~\ref{fig:rq3-network}). Most of these citations (49) point to the other state's AI guidance and 12 to another resource from that state, such as policy templates or acceptable-use policies. 
Most peer citations point readers to another state's work rather than build on it.  Of the 61 citations, 53 (87\%) are references: 36 reference-list or footnote entries and 17 resource listings. The remaining 8 (13\%) are explicit adoptions (purple arrows in Figure~\ref{fig:rq3-network}). For example, North Carolina and Ohio build on Washington's Human--AI--Human model and student AI usage model and North Dakota builds on North Carolina's implementation roadmap. 
Oregon (August 2023), North Carolina (January 2024), and Washington (January 2024) were among the first states to publish a guidance which partly explains why they are also among the most cited documents \cite{Admin_2025}. However, timing alone does not fully explain their position: California published in September 2023 and is cited by only 4 states.  
Overall, peer citations suggest that states do learn from one another, but this learning flows in one direction from a few early publishers and mostly takes the form of references rather than explicit adoption.

\begin{figure*}[t]
  \centering
  \includegraphics[width=\textwidth,keepaspectratio]{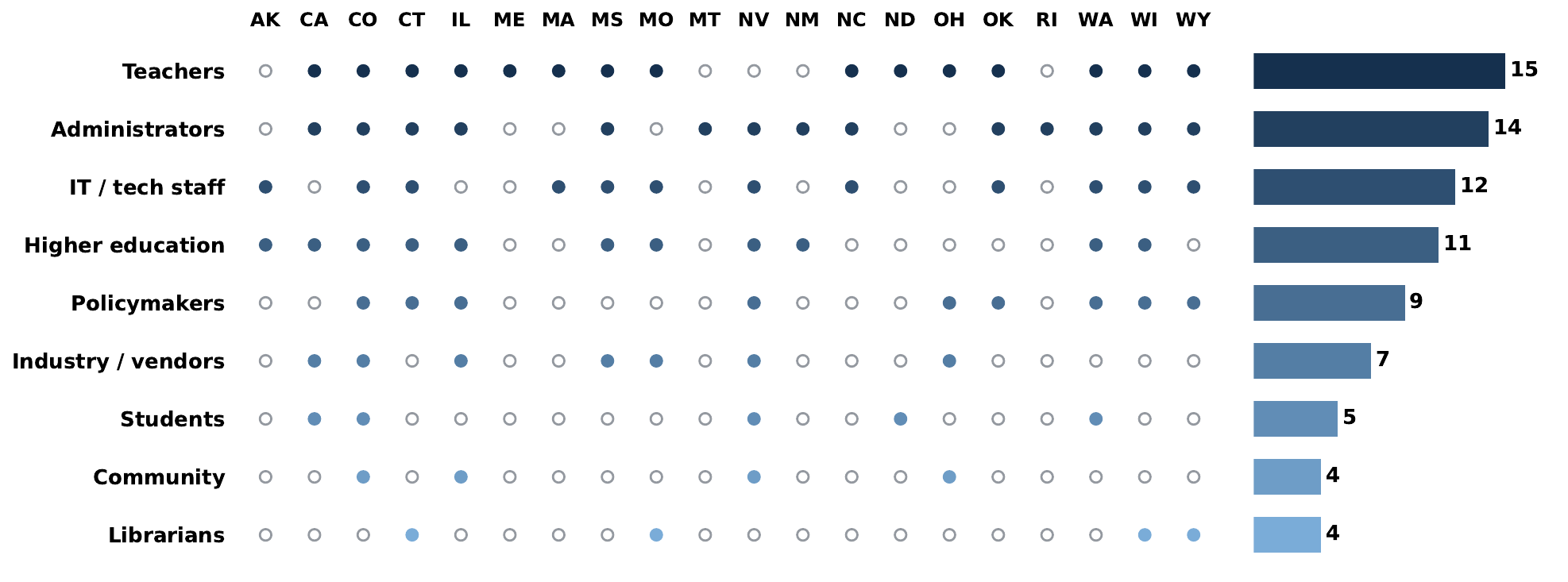}
  \caption{Contributors mentioned in state AI guidance. Bars show the number of states naming each group sorted by frequency. The 20 states naming at least one group are shown alphabetically; 16 states naming none are omitted.}
  \label{fig:rq3-contributors}
\end{figure*}

\subsubsection{\textbf{RQ3C (Contributors): Who is named as a contributor in the guidance documents?}}

Contributors come overwhelmingly from within the education system. The 22 documents with credits name 565 people, 49 organizations, and 30 group statements (Appendix~\ref{app:codebook}). Of the 491 people whose employer is stated, 43\% work for K-12 districts or schools and 28\% for the state department of education, while nonprofits (8\%) and associations (5\%) account for small shares. Outside organizations are thus far more visible as cited sources than as contributors: of the 22 states with credits, 21 cite a nonprofit but only 7 name one among their contributors, and 20 cite a professional association but only 9 name one.
Educators are well represented among contributor groups, but students and families are not. Teachers are named as contributors in 15 states, administrators in 14, and IT or technology staff in 12, whereas students take part in 5 states and no state names parents or families (Figure~\ref{fig:rq3-contributors}), even though 83\% of documents address families as stakeholders (RQ1). Overall, outside expertise enters guidance mainly through citation rather than participation, and the groups who experience AI in the classroom more directly have little or no role in writing the guidance that governs it.

\section{Discussion}

\subsection{States share priorities and epistemic sources}
Federalism is often discussed as allowing states to act as \emph{laboratories}~\cite{USStateLocalGovernments2007}, testing policies and informing federal level action. Beyond government agencies and laws, the most cited sources in guidance documents are advocacy and non-profit groups TeachAI, Code.org, and Digital Promise. These organizations are closely linked; Code.org helped launch TeachAI~\cite{schlosser-2023} and all three coauthored TeachAI's AI toolkit~\footnote{https://www.teachai.org/toolkit}. This suggests that the evidence space for AI guidance may be drawn from narrow, yet prominent, organizations producing reliable evidence accessible and legible to policymakers. Tyler and Gerken argue that ``states as laboratories'' ignores the wide coordinated network of third-party organizations (e.g., non-profits, advocacy groups, funders) that also drive policy innovation \cite{tyler-2022}. Similarly, Salamon has argued that the U.S. is increasingly shifting towards outsourcing government function to third-parties \cite{salamon-2002}. We offer a concrete example of both arguments. Through their AI guidance, states are disseminators of third-party information. While they can still make independent policy choices, we observe that many states are ``soft'' out-sourcing to third-parties: instead of contracting or procuring from them, they are using them for knowledge production and policy innovation.

This observation may also reflect limited resources, time, or technical expertise within state agencies. For example, during remote teaching in 2020 due to COVID-19, most state departments did not have the capacity to guide districts and instead turned to non-governmental resources \cite{willse2024state}. Because guidance occur under resource constraints, agencies may draw on organizations they view as experts. This reliance may explain the overlapping priorities we observe across responsible use \& equity topics. Furthermore, TeachAI's AI toolkit template is used word-for-word in Wyoming's guidance, and is adapted in West Virginia's and Utah's guidance (as can be seen in Appendix Figure~\ref{fig:adoption-verbatim}). To measure this overlap systematically, we compared each state's guidance to TeachAI's guidance using counts of shared word sequences and cosine similarity in sentence-level embeddings (Sentence-BERT~\cite{reimers2019sentence}). This analysis captures diffusion that explicit citation analysis would miss: We find 6,239 shared overlapping 10-word phrases across documents (Appendix Figure~\ref{fig:ngram-copying}) and 88 shared sentences across documents with cosine similarity $\ge.98$ (Appendix Figure ~\ref{fig:cosine-sim}). 

\subsection{Underspecification in accountability and infrastructure}
Teachers surveyed in prior work reported making highly individualized decisions in the absence of clear AI implementation guidelines, and wanted to be co-developers of often top-down policies and standards \cite{dangol2026relief}.
Relative to these needs, we identify two main sources of underspecification: who will actually carry out the guidance and with what infrastructure or resources?  Especially for procuring, implementing, and evaluating AI tools, accountability requires specification of who is tasked with each step. Even Oklahoma's guidance that assigns ``District Technology Teams'' to develop of district AI policy and assessments does not specify which job roles should be held accountable. 
In a similar issue, there is less of an emphasis on the infrastructure (human and technical) needed to actually support this technology. \textit{The Charisma Machine}, describes how utopian visions of technology can overshadow the practicality of the human and technical infrastructure in computer-based education \cite{ames-2019}, highlighting issues in integrating tech into schools without proper infrastructure needed to use it and evaluate its impact. We do find some examples of practical advice in implementing AI as a piece of educational infrastructure, such as Rhode Island's advice for setting contractual not-to-exceed financial amount because of evolving model prices (Rhode Island, p47) and North Carolina's emphasis on investing in \emph{``AI Literacy Training for ALL staff and students FIRST before investing in tools''} (North Carolina, p21). We acknowledge these observations of underspecification can also be due to tensions for policymakers, where highly prescriptive guidance may overburden already taxed school districts or cause guidance to become stale as AI evolves quickly. Less specification can also allow districts and schools to interpret guidance to their own contexts. However, as guidance offers more clarity rather than mandates, and is more adaptable than legislation, it may still be beneficial to be specific about how to achieve guidance goals through actionable steps and update as technology evolves. 

\subsection{Pushing AI as the newest EdTech}
Along with utopian visions from technologists and educators about how technology can, or should, change how we learn, technology companies themselves influence the integration of new educational technology (EdTech) in schools. In ``Coding Kids'' Natasha Singer documents how big tech companies---and the organizations they fund, such as Code.org---have influenced millions of children learn through corporate-branded lessons and tools \cite{singer2026coding}. Recent news reported that Google will turn off search for schools that want to block A.I. entirely \cite{goldstein-2026}. Related to this, we find that states often use a rhetoric of urgency and an ``AI-driven future'' in framing why schools require AI workforce preparation and upskilling. This is also seen in the framing of ``personalized learning'' as a positive reason for adoption, without actual definitions of this use case and evidence as to why. Such framing (whether intentionally or not) in guidance documents mimic earlier 2000's and 2010's push for integration of computers and coding in schools. Yet, the rhetoric of urgency in integrating AI to upskill our future workforce is not present when we compare to the need to build infrastructure or accountability. This may be due less so to corporate influence and more of a government ``soft'' outsourcing of epistemic knowledge. This can, in turn, allow for outside parties to push for AI tools in the classrooms without first asking whether relevant infrastructure or accountability mechanisms are in place. 

\subsection{K-12 AI guidance as one policy instrument of many}
Despite the critiques discussed in this section, guidelines on AI integration in schools are a step in the right direction, especially if they encourage critical thinking about when and how to use AI and how AI can perpetuate social biases. Public policy literature discusses how multiple policy instruments can be used to achieve a goal \cite{salamon-2002, bardach-2023}. Lester Salamon argues that, in fact, the unit of analysis to study in public policy is no longer individual institutions or actors but rather the policy instruments that we use to enact goals \cite{salamon-2002}. The instrument of guidance, while not directly enforceable like statutes or regulations, do promulgate rules and enforce positions held by federal and state administrations \cite{haeder2020policies}. We find that for some states, guidance on education may be able to inform behavior quicker than legislation. We find that five out of the eight states that discuss potential harms of AI companions---West Virginia, Montana, North Carolina, Oklahoma, Illinois---have not yet passed any legislation on the topic of AI companions, despite it being a main concern across the 2025-26 legislative session in the United States \cite{brennen-2026}. Though, just as guidance may provide valuable rules, it may also overlook or omit topics already covered by the state legislature, as many states with laws on AI companion chatbots for minors do not include any consideration for these laws in their guidance. This may be an oversight, a conscious choice, or just due to the fact that states do not regularly update their guidances. This study reinforces that administrative guidance is a utilized policy instrument, though future work is needed to understand which policy instruments may be the most effective in regulating behavior related to AI in schools.  

\subsection{Recommendations for guidances}
We do find comprehensive and actionable state guidances, such as Illinois' thorough review and Ohio's model policy for AI use. Every guidance has important information for schools, however, as we show in our findings and discussion, policies can address more issues with more details. We present five recommendations for updated or future guidance:

\textbf{Requiring annual or regular updates.} We find that guidance documents that have been updated in the past year tend to contain more guidance on novel affordances from AI (e.g., AI companions). We also faced difficulty in understanding the versioning history of many documents, sometimes unable to pinpoint when the guidance was even released or updated. To address this, we recommend annual or regular updates to guidance documents, where versioning history is included in the document or in the same source as to where the public can access the guidance document. Examples of versioning history can be found in North Carolina (p3) and Missouri (p3) guidance documents. 

\textbf{Improved participation of stakeholders.} We identify three places to improve participation of stakeholders. The first is in authorship of guidelines. Only five guidances involve students in writing and none include parents. Additionally, teacher unions are not considered as stakeholders in the guidance documents nor are they identified as authors. Yet, two large unions---National Education Association and the American Federation of Teachers---both argue in their AI guidances that unions should be included in setting and implementing AI policy \cite{NEA, AFT}. We suggest including teachers unions, students, and parents when drafting state and local guidance and policy. The second is to include parents through providing consent forms about the AI their child will be using, as seen in West Virginia's privacy policy consent forms for parents (p.29). Third, previous research has shown that teachers want ongoing institutional support for AI literacy and express a desire for collaborative learning environments \cite{bautista-2025}. To address this, guidance can encourage the participation of teachers in more specific, ongoing, and compensated workshops and collaborative learning.

\textbf{Providing clear roles and tasks.} Accountability can also be improved by providing clear roles and tasks for existing roles or directions for the development of new roles to address unique needs. This is seen in Alabama's guidance that provides a template for Local Education Agencies (LEA) to fill in: \emph{``The (replace with the name of LEA) is committed to maintaining a risk register that will record AI systems: Not in compliance with the NIST framework, federal, state and laws, rules, and regulations...Person names and roles who are accountable and responsible for mitigating the risks.''} (Alabama, p9).

\textbf{More consideration for why AI should be used.} We find that many arguments for the usage of AI focus on potential benefits of tools such as personalized learning or use a sense of urgency to describe the need to prepare students for a new, quickly changing AI workforce. Such language can make AI use appear inevitable. Instead, guidance documents should provide clear definitions as to what AI tools are available, how they have had positive impact, and how they can be implemented. 

\textbf{Demographic and resource responsiveness.} We do not find many patterns between the topics covered in guidance and the demographics of a state, with only the mention of tribal/indigenous communities being associated with having a higher indigenous enrollment in schools. States vary in their demographic and infrastructural capacities, yet high reliance on resources from third-party organizations like TeachAI, Code.org, and Digital Promise can potentially collapse those differences, and make guidance documents less culturally specific. We suggest that states should address specific community needs, and should especially consider the infrastructure and resources of many of their districts in their ability to implement the AI guidance.

\section{Limitations and Future Work}
Our work has several limitations. In terms of data, our analysis is limited to the 36 jurisdictions whose state departments have issued guidance. Our document coding process is binary, and does not capture the depth, specificity, or attitude towards each topic. We also only rely on latest issued documents by each jurisdiction without accounting for how these documents may have evolved over time. We believe there is an opportunity for future work to study the evolution of AI policy and guidance across states. Furthermore, because documents have been issued at different points over roughly three years, earlier documents, such as Washington and Oregon, have had more time to be cited, while more recently updated documents like Montana's have had the advantage to include emerging issues like AI companionship. We report these patterns descriptively without adjusting for time frame. 
We leave the exploration of the mechanisms behind policy diffusion for future work to explore. Finally, state guidance documents are only one of the instruments affecting policy and governance. Future work should explore the relationship between guidance and other policy documents such as statutes, and teacher union statements to compare whether concerns raised in these documents, such as around labor, are reflected in guidance priorities. Along these lines, future work should also investigate potential impacts of state guidances on actual AI implementation in classrooms. 

\section{Conclusion}
In this study we review 36 state guidance documents for AI in K-12 education. We find that issues around academic integrity, data privacy, and algorithmic bias appear across most documents, while budgeting, actionable roadmaps, and specific underserved populations receive far less attention. Guidance is mostly directed towards school districts and educators, whereas students and IT staff are rarely given direct guidance. The priorities in guidance documents show little relationship to state demographics and administrative capacities. Instead, states rely on a shared set of epistemic sources, including laws, government agencies, and advocacy and non-profit groups such as TeachAI and Code.org. States also cite each other, with peer citations concentrated on a few early states like Washington and Oregon. Rather than acting as independent \emph{laboratories of democracy}, we find states to be quite \emph{interdependent}. This allows states to learn from one another in a quickly changing landscape of technology shaped by AI, but also concentrates influence over how million of students encounter technology in the classroom. 
Our findings and discussion offer insight into state guidances' as policy instruments more broadly, and highlight the need for greater transparency about who shapes state AI guidance, broader representation of affected communities, and more actionable recommendations that respond to the diverse needs of schools and students. 

\section*{Generative AI Disclosure Statement}

Generative AI was used for light editing in the writing of the draft (grammar and rephrasing). Tools used include GPT-6 and Claude Opus 5.5. 
Claude Sonnet 5 was used within research workflow for validation of content analysis as described in Section~\ref{sec:content_analysis}. We also use Claude Code (Opus 4.8) as a general purpose coding assistant for creating visualizations, formatting of tables and for assisting with analyses code for RQ2, RQ3, and text-reuse analysis (Section 5.1, Appendix~\ref{fig:cosine-sim}, Appendix~\ref{fig:ngram-copying}). Claude Opus 4.8 drafted the initial source dictionary mapping sources to sectors, as described in Section~\ref{sec:rq3-methods}; one author reviewed every entry and mapping. Authors assume full responsibility for the content produced in relation to this work. 

\section*{Ethical Considerations Statement}
This study analyzes publicly available guidance documents and does not involve human participants, interventions, or private data. A subset of guidance documents name individuals that served as contributors to the document; we report contributor-related analyses in aggregate, and do not name individuals. For our content analysis, we use an LLM-based system as a secondary check in content coding; every proposed change is reviewed by two human coders. In our citation and contributor analysis, we use an LLM-based system for dictionary creation which is reviewed by one of the authors. Our analysis is descriptive and is not meant to rank states; we highlight examples of good practices and gaps, realizing that agencies responsible for the creation of these documents are under severe time and resource constraints. Our findings on text similarity and reuse are description of patterns of reliance, and not meant to suggest misconduct by any state or agency. 

\section{Acknowledgments}
We thank members of the DSCO group at the University of Washington Information School, especially Belén Saldías, Soham De, Navreet Kaur, Ott Toomet, and Yongxu Sun, for productive discussions that shaped the initial ideas in this paper. We are also grateful to Aayushi Dangol for discussions that informed the early development of this work, and to Sarah Tran for pointing us to key references for the citation analysis.

\section{Other Disclosures}
The first author (SAM) was a member of Washington Office of Superintendent of Public Instruction (OSPI)'s AI Advisory Group and coauthored the \href{https://ospi.k12.wa.us/sites/default/files/2024-07/ai-guidance_foundations.pdf}{AI guidance for the state of Washington}. 

\bibliographystyle{ACM-Reference-Format}
\bibliography{sample-base}

\newpage
\appendix

\section{Appendices}

\subsection{Coding for RQ1}
\label{sec:rq1_coding}
To identify guidance priorities and audiences (RQ1), our analysis followed a three-stage content analysis process.  First, two authors reviewed an initial subset of 3 state documents using exploratory open coding to identify potentially relevant themes and issues. These preliminary observations were then discussed and used to develop and refine a set of 30 analytic questions (see Appendix Table~\ref{tab:coding-legend}). In the second stage, the resulting content analysis framework was subsequently applied deductively across the full document corpus of 36 guidance documents. For this stage, two authors independently coded the documents across 30 questions (i.e. 1,080 cells), taking informal notes throughout. The human coders were concordant on 86.3\% (n=932) of cells. The remaining disagreements were resolved through discussion. In the third stage a coding pass was conducted using Claude Sonnet 5 as an additional consistency check. Claude's independent codes were concordant with the final agreed upon set of content analysis from the two-coders on 86.6\% (n=935) of the cells. The discrepancies or potential issues highlighted by Claude were subsequently reviewed and used to prompt discussion between the two human coders to determine if anything was missed before final coding decisions were made. Adjudication upheld Claude's code decision in 41.4\% (n=60) of the highlighted cases, identifying genuine errors in human consensus. 58.6\% (n=85) of the rest of the codes remained unchanged. We used Claude cautiously as our questions were binary and required little interpretation. This was an important step as our work relies on whether or not specific topics are discussed across states, and we did not want to inadvertently miss any important details or make claims that were untrue. For example, both the coders initially coded \emph{No} for the question on \emph{``Are deepfakes or synthetic media explicitly discussed?''} for the state of Delaware. However, Claude marked it as \emph{Yes}, and flagged a viable quote from the document for us to review: \emph{``Bullying and harassment by using AI tools to manipulate media in order to impersonate others can have severe consequences for students' well-being.''}

\definecolor{FamMeta}{HTML}{FCD5CE}
\definecolor{FamStruct}{HTML}{D8E2DC}
\definecolor{FamAud}{HTML}{F7E1D7}
\definecolor{FamPed}{HTML}{FFD7BA}
\definecolor{FamAct}{HTML}{BEE1E6}
\definecolor{FamWork}{HTML}{FFE5EC}
\definecolor{FamResp}{HTML}{CDDAFD}
\providecommand{\famsw}[1]{\raisebox{-0.15ex}{\setlength{\fboxsep}{0pt}\fcolorbox{black!35}{#1}{\rule{0pt}{1.3ex}\rule{1.3ex}{0pt}}}}
\begin{table}[htbp]
\centering\small
\caption{Shorthand labels used in Table~\ref{tab:coding-grid}, grouped into seven families, with the full coding question for each (as worded in the coding sheet).}
\label{tab:coding-legend}
\begin{tabular}{@{}l p{0.66\linewidth}@{}}
\toprule
Shorthand & Full coding question \\
\midrule
\rowcolor{FamMeta}\multicolumn{2}{@{}l}{\textbf{Document metadata}} \\
\rowcolor{FamMeta!15}Version history & Does the document have explicit version numbering or a revision history? \\
\rowcolor{FamMeta!15}Year (recent) & What year was the most recent version published? \\
\rowcolor{FamMeta!15}Month (recent) & What month was the most recent version published? \\
\rowcolor{FamMeta!15}AI self-disclosure & Does the document disclose that AI tools were used in its own creation? \\
\midrule
\rowcolor{FamStruct}\multicolumn{2}{@{}l}{\textbf{Document structure}} \\
\rowcolor{FamStruct!15}Guiding principles & Does the document contain a section for guiding principles? \\
\rowcolor{FamStruct!15}Glossary / definitions & Does the document include a definition or glossary section for key AI terms? \\
\rowcolor{FamStruct!15}Review cycle & Does the document commit to a periodic (e.g., annual) review and update cycle? \\
\midrule
\rowcolor{FamAud}\multicolumn{2}{@{}l}{\textbf{Audience}} \\
\rowcolor{FamAud!15}Administrators & Is guidance given directly to school administration? \\
\rowcolor{FamAud!15}Teachers & Is guidance given directly to teachers? \\
\rowcolor{FamAud!15}Students & Is guidance given directly to students? \\
\rowcolor{FamAud!15}Families / parents & Are families and parents mentioned as stakeholders or given guidance? \\
\rowcolor{FamAud!15}IT / tech staff & Is guidance given directly to IT / technology staff? \\
\midrule
\rowcolor{FamPed}\multicolumn{2}{@{}l}{\textbf{Pedagogy \& instruction}} \\
\rowcolor{FamPed!15}Professional development & Does the document include explicit professional development (PD) guidance for educators? \\
\rowcolor{FamPed!15}Assessment redesign & Does the document provide guidance on redesigning curriculum (or assessments) in light of AI? \\
\rowcolor{FamPed!15}Personalized learning & Is AI-enabled personalized or adaptive learning cited as an educational use case? \\
\midrule
\rowcolor{FamAct}\multicolumn{2}{@{}l}{\textbf{Actionability}} \\
\rowcolor{FamAct!15}Roadmap / timeline & Does the document include an actionable plan, timeline, or roadmap? \\
\rowcolor{FamAct!15}Procurement / implementation & Does the document include specific instructions on actionable steps to get AI in the classrooms (e.g. procurement, implementation)? \\
\rowcolor{FamAct!15}Funding / budget & Does the document mention funding, budget allocations, or resource requirements for AI implementation? \\
\rowcolor{FamAct!15}Vendor evaluation & Does the document provide guidance on evaluating or approving AI tool vendors? \\
\midrule
\rowcolor{FamWork}\multicolumn{2}{@{}l}{\textbf{Workforce \& career readiness}} \\
\rowcolor{FamWork!15}Workforce / jobs / upskilling & Is workforce preparation, impact of AI on jobs or upskilling mentioned? \\
\midrule
\rowcolor{FamResp}\multicolumn{2}{@{}l}{\textbf{Responsible use \& equity}} \\
\rowcolor{FamResp!15}Deepfakes / synthetic media & Are deepfakes or synthetic media explicitly discussed? \\
\rowcolor{FamResp!15}Equity / digital divide & Is equity of access mentioned? \\
\rowcolor{FamResp!15}Algorithmic bias & Is algorithmic bias or algorithmic fairness in AI systems specifically addressed? \\
\rowcolor{FamResp!15}English language learners & Are English language learners (ELL) or multilingual learners specifically addressed as a student population? \\
\rowcolor{FamResp!15}Rural / underserved & Are rural or underserved districts explicitly addressed as a special implementation context? \\
\rowcolor{FamResp!15}Tribal / Indigenous & Is the impact of AI on indigenous or tribal communities explicitly addressed? \\
\rowcolor{FamResp!15}Accessibility \& disability & Are students with disabilities explicitly considered (e.g., IDEA, Section 504, ADA)? \\
\rowcolor{FamResp!15}Academic integrity & Is academic integrity, plagiarism, or source attribution explicitly addressed? \\
\rowcolor{FamResp!15}Data privacy \& governance & Are data privacy and governance mentioned as a concern? \\
\rowcolor{FamResp!15}AI companion apps & Are AI companion apps or AI companionship mentioned? \\
\bottomrule
\end{tabular}
\end{table}

\newpage

\subsection{Data sources for RQ2}

In the following table, we highlight the data sources used for RQ2. 

\begin{longtable}{
    >{\RaggedRight\arraybackslash}p{0.30\textwidth}
    >{\RaggedRight\arraybackslash}p{0.25\textwidth}
    >{\centering\arraybackslash}p{0.10\textwidth}
    >{\RaggedRight\arraybackslash}p{0.25\textwidth}
}
\caption{Covariates, Data Sources, and Corresponding Guidance Dimensions}

\label{tab:rq2-data-sources} \\

\toprule
\textbf{Covariate} &
\textbf{Data Source} &
\textbf{Year} &
\textbf{Dimension in Guidance} \\
\midrule
\endfirsthead

\multicolumn{4}{l}{\textit{Table \thetable{} continued}} \\
\toprule
\textbf{Covariate} &
\textbf{Data Source} &
\textbf{Year} &
\textbf{Dimension in Guidance} \\
\midrule
\endhead

\bottomrule
\endfoot

\multicolumn{4}{c}{\textbf{Demographic Characteristics}} \\
\midrule

Percentage of rural population &
\href{https://data.census.gov/table/DECENNIALDHC2020.P2}
{U.S. Census Bureau} &
2020 &
Rural/underserved \\

\addlinespace

Percentage of English language learners &
\href{https://nces.ed.gov/programs/digest/d23/tables/dt23_204.20.asp}
{NCES Digest 2023, Table 204.20} &
2021 &
English language learners \\

\addlinespace

Percentage of American Indian and Alaska Native (AI/AN) &
\href{https://nces.ed.gov/programs/digest/d23/tables/dt23_203.70.asp}
{NCES Digest 2023, Table 203.70} &
2022 &
Tribal/Indigenous \\

\addlinespace

Percentage of children served under the Individuals with
Disabilities Education Act (IDEA) &
\href{https://nces.ed.gov/programs/digest/d23/tables/dt23_204.70.asp}
{NCES Digest 2023, Table 204.70} &
2022--2023 &
Accessibility \& disability \\

\midrule
\multicolumn{4}{c}{\textbf{State Capacity and Socioeconomics}} \\
\midrule

Percentage of public school students eligible for free
or reduced-price lunch &
\href{https://nces.ed.gov/programs/digest/d23/tables/dt23_204.10.asp}
{NCES Digest 2023, Table 204.10} &
2022--2023 &
Equity/digital divide \\

\addlinespace

Percent of households in the state that have any
broadband subscription &
\href{https://data.census.gov/table/ACSST1Y2023.S2801?q=S2801\&g=010XX00US\%240400000}
{U.S. Census Bureau, S2801\_C02\_014E} &
2023 &
Equity/digital divide \\

\addlinespace

Per-pupil spending &
\href{https://www.census.gov/programs-surveys/school-finances/data/tables.html}
{U.S. Census Bureau, Annual Survey of School System Finances} &
2024 &
Professional development,
funding/budget, upskilling \\

\midrule
\multicolumn{4}{c}{\textbf{Political Leaning}} \\
\midrule

Partisan composition of state legislatures nationwide
by percent &
\href{https://ballotpedia.org/Partisan_composition_of_state_legislatures}
{Ballotpedia} &
2026 &
Responsible use and equity \\

\end{longtable}

\newpage

\subsection{Additional methods and results for RQ3}
\label{app:rq3-sources}
This appendix gives the details of the procedures described in Section~\ref{sec:rq3-methods}, followed by additional results. Table~\ref{tab:rq3-entities} summarizes the three kinds of information we extract. Every match is stored with its page and surrounding text, so each citation can be traced to a passage in a document.

\begin{table*}[h]
\centering
\small
\setlength{\tabcolsep}{4pt}
\begin{tabularx}{\textwidth}{
  >{\raggedright\arraybackslash\hsize=0.68\hsize}X
  >{\raggedright\arraybackslash\hsize=1.32\hsize}X
  >{\raggedright\arraybackslash\hsize=1.27\hsize}X
  >{\raggedright\arraybackslash\hsize=0.73\hsize}X}
\toprule
\textbf{Information} & \textbf{Method} & \textbf{Example} & \textbf{Final data} \\
\midrule
Cited sources & Hyperlinks resolved to the organization behind each website; text matched against a dictionary of organizations, frameworks, and laws; each source assigned one of ten types & \texttt{studentprivacy.ed.gov} $\rightarrow$ Student Privacy Policy Office (federal government); ``FERPA'' $\rightarrow$ Family Educational Rights and Privacy Act (law) & 898 state-source pairs; 328 sources \\
Citations of other states & Search for state names and department acronyms; each passage read and coded by one author for how it is used & ``borrowed from'' Washington (Wisconsin) $\rightarrow$ explicit adoption & 61 citations among 21 states \\
Contributors & Credits and acknowledgments read; each entry coded by one author for affiliation and contributor group & ``a writing team made up of educators, students, and stakeholders'' $\rightarrow$ group statement (teachers, students) & 565 people; 49 organizations; 30 group statements \\
\bottomrule
\end{tabularx}
\caption{Information extracted from the 36 guidance documents for RQ3, with the method used, an example, and the size of the final data.}
\label{tab:rq3-entities}
\end{table*}

\subsubsection{Document Preprocessing}
From the 36 documents described in Section~\ref{sec:data-collection-guidances}, we extract the text of each page and a list of the document's hyperlinks.  For PDF files, we extract both with PyMuPDF \cite{pymupdf}; for Word files, we parse them from their XML. Since Maine's guidance consists of slide images, we used Tesseract OCR \cite{smith2007tesseract} to obtain its text. 

\subsubsection{Identifying Cited Sources}
Since a document can refer to a source by linking to it, naming it in text, or both, we run three searches over the extracted text. The first collects every hyperlink and printed web address and maps its website to the organization behind it; for example, \texttt{studentprivacy.ed.gov} maps to the Student Privacy Policy Office of the U.S.\ Department of Education. The second searches the text for the names and abbreviations of organizations, frameworks, and laws, such as FERPA; it is the only search that finds laws, which have no website. The third searches for the names of other states. Table~\ref{tab:rq3-channels} reports the number of matches each search returns. The searches run on the same text but look for different things, so each finds sources the others miss: of the final state-source pairs, 35.5\% were found only through a link, 28.8\% only through a name, 29.6\% through both, and 6.0\% through a mention of another state.

\begin{table}[h]
\centering
\small
\begin{tabular}{p{2.6cm}p{7.2cm}r}
\toprule
\textbf{Search} & \textbf{What is searched} & \textbf{Matches} \\
\midrule
Hyperlinks & Clickable links and printed web addresses, matched by host name & 3{,}993 \\
Names in the text & Names and abbreviations of the organizations, frameworks, and laws in the dictionary & 2{,}556 \\
Other states & State names and department acronyms near terms such as ``Department'' or ``guidance'' & 189 \\
\midrule
Total & & 6{,}738 \\
\bottomrule
\end{tabular}
\caption{Matches returned by each search, before removing matches that do not refer to outside sources.}
\label{tab:rq3-channels}
\end{table}

\emph{The dictionary.} Linking citations to sources relies on a dictionary that lists, for each known source, its websites, the names and abbreviations under which it appears in text,and a sector label (e.g., nonprofit, federal, law). Table~\ref{tab:rq3-dictionary} shows examples. We drafted the dictionary with the help of Claude Opus 4.8, validated by one author. We then searched the corpus for sources the draft had missed and added them. The final dictionary has 122 entries: 112 are sources, and 10 describe AI products, file hosts, and social media, which we use only to find matches to remove.

\begin{table*}[h]
\centering
\small
\begin{tabularx}{\textwidth}{@{}l l >{\raggedright\arraybackslash}X >{\raggedright\arraybackslash}X@{}}
\toprule
\textbf{Source} & \textbf{Sector} & \textbf{Websites} & \textbf{Names in text} \\
\midrule
TeachAI & Nonprofit & \texttt{teachai.org} & ``TeachAI'', ``AI Guidance for Schools Toolkit'' \\
U.S.\ Dept.\ of Education & Federal & \texttt{ed.gov}, \texttt{tech.ed.gov}, \ldots & ``U.S. Department of Education'', ``Office of Educational Technology'' \\
FERPA & Law & none & ``FERPA'', ``Family Educational Rights and Privacy'' \\
News and magazines & Media & \texttt{forbes.com}, \texttt{bbc.com}, \ldots & none \\
ChatGPT (to remove) & AI product & \texttt{chatgpt.com} & ``ChatGPT'' \\
\bottomrule
\end{tabularx}
\caption{Example dictionary entries. A source with no website can be found only by name, and one with no name only by link.}
\label{tab:rq3-dictionary}
\end{table*}

\emph{Hyperlinks.} A website is matched to the dictionary entry with the longest matching ending. This lets an office within an agency be recognized separately from the agency itself: \texttt{studentprivacy.ed.gov} resolves to the Student Privacy Policy Office, whereas \texttt{ed.gov} and its other subdomains resolve to the U.S.\ Department of Education. A website that is not in the dictionary is attributed to a state department of education or a school district when the address identifies one, and otherwise to its own domain.

\emph{Names in the text.} We join lines and collapse repeated spaces, as names in PDF text are often broken across lines. Company names pose a particular problem, since a document that mentions Google or Microsoft is usually naming a product rather than citing the company. We therefore count a company name only when a word such as \emph{report}, \emph{framework}, \emph{guidance}, \emph{policy}, or \emph{initiative} appears within 70 characters of it.

\emph{Other states.} We search for each state's full name and postal abbreviation and for the acronyms of six state departments of education (OSPI, NCDPI, WVDE, USBE, ALSDE, and OSDE). Several state names also occur in other names, so we exclude ``West Virginia'' from matches for Virginia, require an adjacent term such as ``State'' or ``OSPI'' for Washington, and skip names such as ``Washington Post'' and ``University of \ldots.''

\emph{Removing and merging matches.} We remove 1{,}387 of the 6{,}738 matches (21\%): 811 links to the issuing state's own website, 334 links to file hosts and social media, and 242 mentions of AI products named as examples of tools, such as ChatGPT or Canva. We also set aside 78 matches to commercial websites that are not in the dictionary and exclude 14 pairs that refer to local school districts. We then merge the remaining matches into one entry per state and source. Merging by organization rather than by state name prevents organizations whose names include a state from being counted as that state's guidance; for example, \texttt{edpolicyinca.org} resolves to Policy Analysis for California Education (PACE), not to California. Our unit of analysis is the state-source pair: a state cites a source if its guidance mentions it at least once. We do not count repeated mentions, since document length varies widely. 

\subsubsection{Source Types and Validation}
Table~\ref{tab:rq3-types} lists the ten source types with examples and counts. Each dictionary entry's sector label maps to one of the ten types (e.g., \emph{federal} to federal government), and a few sources, such as membership associations, are typed individually. Websites outside the dictionary are typed by their address ending: \texttt{.edu} as a research institute or university, \texttt{.gov} and \texttt{.mil} as the federal government, foreign government domains as international organizations, and \texttt{.org} and \texttt{.net} as nonprofits. 

Two coding rules apply while classifying citations. First, each source is typed by the organization that publishes it rather than by the kind of document linked, so a research brief from the College Board is typed as the College Board. Second, membership associations of educators, school boards, principals, and administrators, such as ISTE, CoSN, NSBA, and CABE, are typed as professional associations even when they also engage in advocacy.

\begin{table}[h]
\centering
\small
\begin{tabular}{p{4.0cm}p{4.2cm}rr}
\toprule
\textbf{Type} & \textbf{Examples} & \textbf{Sources} & \textbf{Pairs} \\
\midrule
Advocacy group, nonprofit or union & TeachAI, Code.org, Common Sense Media & 102 & 224 \\
Law or legal requirement$^{\ast}$ & FERPA, COPPA, IDEA & 11 & 128 \\
Research institute, think tank or university & Brookings, Stanford HAI, CIRCLS & 78 & 127 \\
Federal government & U.S.\ Dept.\ of Education, FTC, NIST & 32 & 121 \\
Professional association & ISTE, CoSN, CSTA & 16 & 64 \\
Other state government & Washington, Oregon, North Carolina & 18 & 61 \\
Company or consultancy & Google, Microsoft, OpenAI & 15 & 58 \\
International organization & UNESCO, OECD, European Union & 16 & 58 \\
Media, reference or other$^{\ast}$ & EdWeek, Common Core & 38 & 52 \\
Philanthropic foundation & Kapor Foundation, Walton Family Foundation & 2 & 5 \\
\midrule
Total & & 328 & 898 \\
\bottomrule
\end{tabular}
\caption{Source types, with examples and the number of sources and state-source pairs of each type. $^{\ast}$Types added by us; the others follow the kinds of actors identified in prior work on policy transfer. Local school districts were coded as an eleventh type and are excluded (14 pairs).}
\label{tab:rq3-types}
\end{table}

The validation sample of 150 state-source pairs is stratified by how each pair was found, so that each search is checked: 50 pairs found only through a link, 30 through both a link and a name, 40 only through a name, and 30 through a mention of another state. For each pair, the author read the stored passage and answered four questions: whether the passage refers to the assigned source, whether the source's name is correct, whether its type is correct, and whether the document uses the source in support of its content. 

\subsubsection{Comparing States}
\label{app:rq3-compare}
To compare the sources of two states, we use Jaccard similarity \cite{jaccard1901etude}, the number of sources both states cite divided by the number cited by either. Since the denominator grows with the length of both source lists, a state that cites many sources does not appear similar to every other state merely by citing more. We compute the measure for all 630 pairs of the 36 jurisdictions and report its mean. To test whether neighboring states cite more similar sources, we assign states to the four U.S.\ Census Bureau regions, excluding Puerto Rico, which belongs to none, and compare the mean similarity of pairs in the same region with that of pairs in different regions. We shuffle the region labels across states 2{,}000 times and recompute the difference each time; the one-sided $p$-value is the share of shuffles, counting the observed assignment, whose difference is at least as large as the observed one.

\subsubsection{Coding Peer Citations}
Citations of other states provide the most direct evidence of whether states learn from one another, so one author read the passages for each of the 65 candidate pairs of citing and cited states and coded how the citation is used: an entry in a reference list, an item in a list of resources, an adaptation of the other state's material, a specific idea credited to the state, a recommendation to follow it, a comparison, and an acknowledgment. Only the first three occurred, so we report two categories: \emph{explicit adoption}, which comprises adaptations, and \emph{reference}, which comprises reference-list entries and resource listings. To help the coder find adoptions, the form highlighted phrases such as ``adapted from,'' ``borrowed from,'' ``based on,'' ``modeled on,'' and ``building on'' near the other state's name; the coder read every passage before assigning a code.

Four of the 65 candidates were removed. Three were false matches: Minnesota names California and Virginia only when describing a Digital Promise review of state guidance, and Ohio names California only to note that a roadmap was written for California schools. The fourth cites a Pennsylvania law rather than guidance. On a second reading, we recoded two citations from reference to explicit adoption, because the citing text states that material was ``borrowed from'' (Wisconsin citing Washington) or ``As adapted from'' (Wyoming citing Washington). The 61 citations comprise 8 explicit adoptions, 36 reference-list or footnote entries, and 17 resource listings; 49 cite the other state's AI guidance and 12 another resource from that state. We represent the 61 citations as a directed network from citing to cited state \cite{hagberg2008networkx}.

\subsubsection{Coding Contributors}
\label{app:codebook}
For each document, we read the acknowledgments, credits, introductory pages, and any lists of committee or task force members, and recorded three kinds of entries: people named as authors, committee members, reviewers, or officials; organizations credited with preparing or contributing to the guidance; and group statements, such as ``a writing team made up of educators, students, and stakeholders.'' We relied only on what each document states. We built the list of contributors in three steps. First, a script searched for credit headings, such as acknowledgments, contributors, task force, advisory committee, steering committee, writing team, and working group, in the first ten and last eight pages of each document, or in all pages of documents of four pages or fewer. Second, one author checked each of the 393 person entries from this step against the page and changed the proposed category for 145 of them. Third, we read every credits page again and rebuilt the list, extracting the text with \texttt{pdftotext} in layout mode \cite{poppler} so that names, titles, and organizations stay aligned. The rebuild added 172 people, 110 of them from Colorado, and removed 95 duplicates and 10 entries that were not people, giving the final list of 565 people, 49 organizations, and 30 group statements.

Table~\ref{tab:rq3-codebook} defines the categories coded. A contributor group is assigned from a person's title or a group's name, so a person listed without a title receives no group. A person whose affiliation is not printed is coded as ``other or cannot tell''; this applies to 74 people, including all 13 people named in Maine and 36 of the 51 named in Wisconsin. Of the 93 people whose affiliation we flagged as uncertain, these 74 make up most; the other 19 have a stated role or organization that does not settle the category, and we report them under the category we assigned (11 nonprofit, 4 company or vendor, 3 association, and 1 other government). North Dakota, Ohio, and Wyoming describe their contributors only through group statements. Since 14 of the 36 documents contain no credits, all contributor counts are lower bounds.

\begin{table}[h]
\centering
\small
\begin{tabular}{p{3.2cm}p{9.5cm}}
\toprule
\multicolumn{2}{l}{\textbf{Affiliation (the organization printed with the person's name)}} \\
\midrule
State department of education & A state department of education, state board, or similar state agency \\
K-12 district or school & A school district, school, or regional education service center \\
Higher education & A university or college \\
Nonprofit & A nonprofit, foundation, coalition or advocacy group \\
Association & A professional association, union or membership group \\
Company or vendor & A company, vendor or consultancy \\
Other government & Governor's office, legislature, or another agency \\
Other or cannot tell & No affiliation given, or none of the above \\
\midrule
\multicolumn{2}{l}{\textbf{Contributor group (the role the person or group plays)}} \\
\midrule
Teachers & Classroom teachers and educators \\
Administrators & Principals, superintendents, district or school leaders \\
Higher education & Professors, university or college staff \\
Policymakers & Legislators, board members, commissioners \\
Students & Students or youth \\
Industry or vendors & Companies, vendors, industry \\
Parents or families & Parents, families, guardians \\
IT or technology staff & Technology, IT, data or digital-learning staff \\
Community & Community members, civic or business groups \\
Librarians & Librarians and media specialists \\
\bottomrule
\end{tabular}
\caption{Codebook for contributors. Each person has one affiliation and, when a title is given, one main contributor group.}
\label{tab:rq3-codebook}
\end{table}

\subsubsection{Laws and Frameworks Cited}
\label{app:rq3-extra}
\begin{figure}[h]
  \centering
  \includegraphics[width=\columnwidth,keepaspectratio]{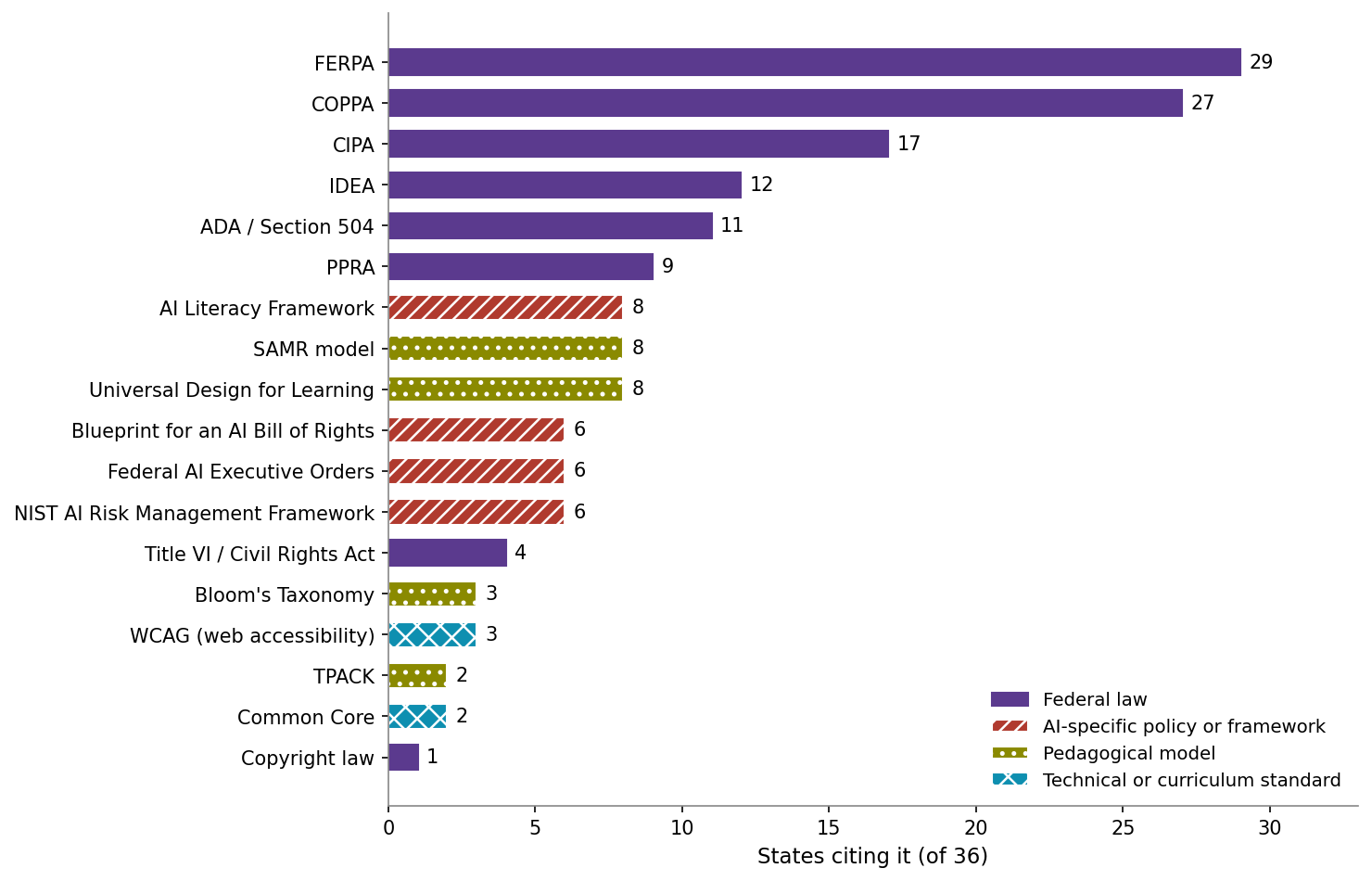}
  \caption{Laws, AI-specific policies and frameworks, teaching models, and standards, by the number of states citing each.}
  \label{fig:laws}
\end{figure}

\begin{figure}[h]
  \centering
  \includegraphics[width=\columnwidth,keepaspectratio]{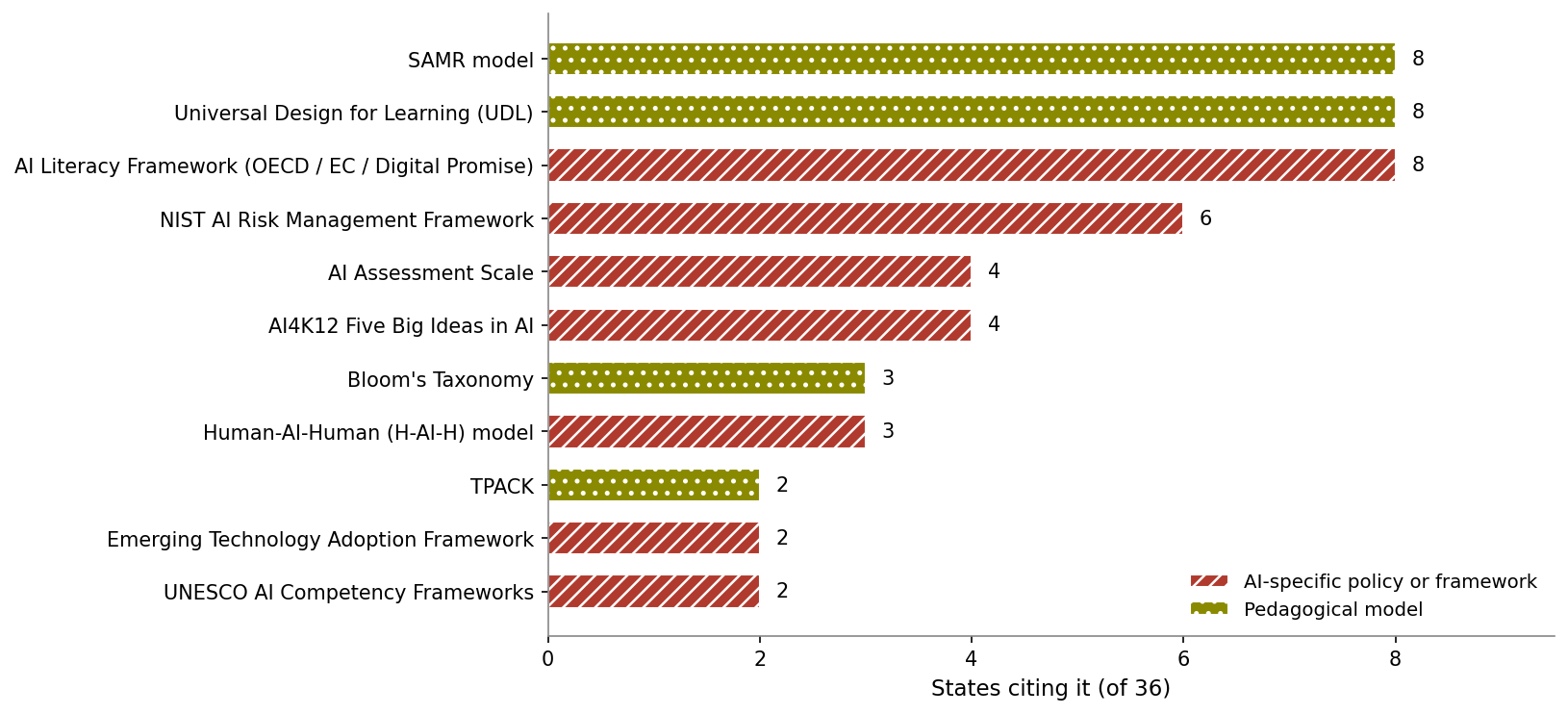}
  \caption{AI-specific frameworks and teaching models cited by two or more states.}
  \label{fig:frameworks}
\end{figure}

Figure~\ref{fig:laws} shows how many states cite each law, AI-specific policy or framework, teaching model, and technical or curriculum standard named in the guidance. Figure~\ref{fig:frameworks} shows the AI-specific frameworks and teaching models, including several that do not appear in Figure~\ref{fig:laws}.

\clearpage

\subsection{Examples of policy adoption and copying across documents}
\label{sec:consine-analysis}

\begin{figure*}[!ht]
  \centering
\includegraphics[width=\textwidth,keepaspectratio]{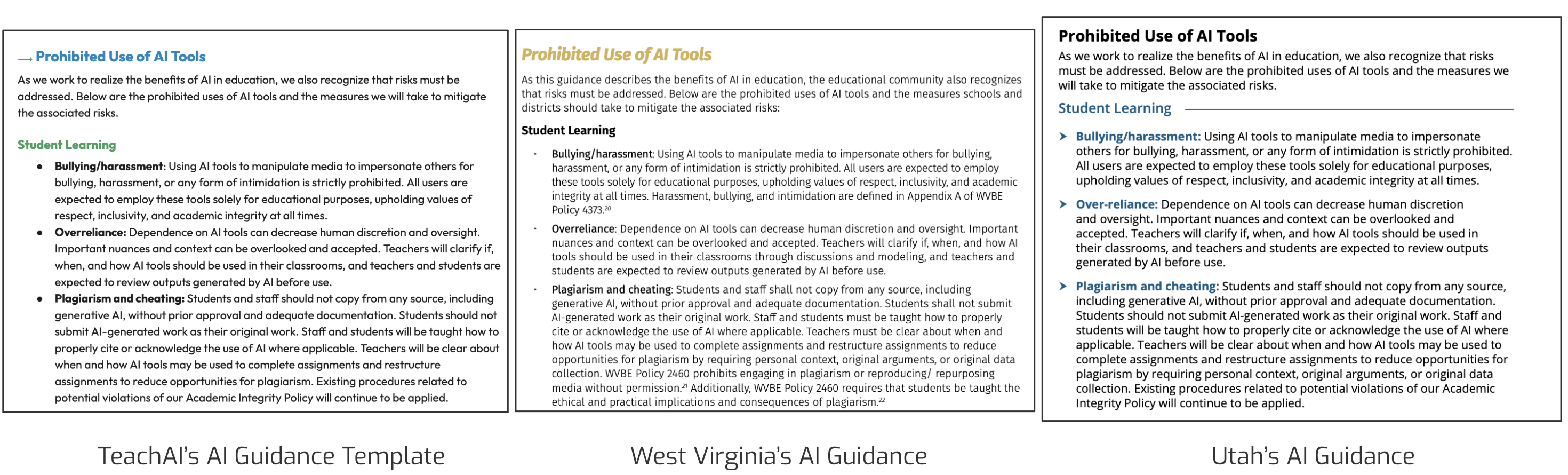}
  \caption{An example of copying between guidance documents of TeachAI, West Virginia, and Utah}
  \label{fig:adoption-verbatim}
\end{figure*}


\begin{figure*}[!ht]
  \centering
  \includegraphics[width=\textwidth]{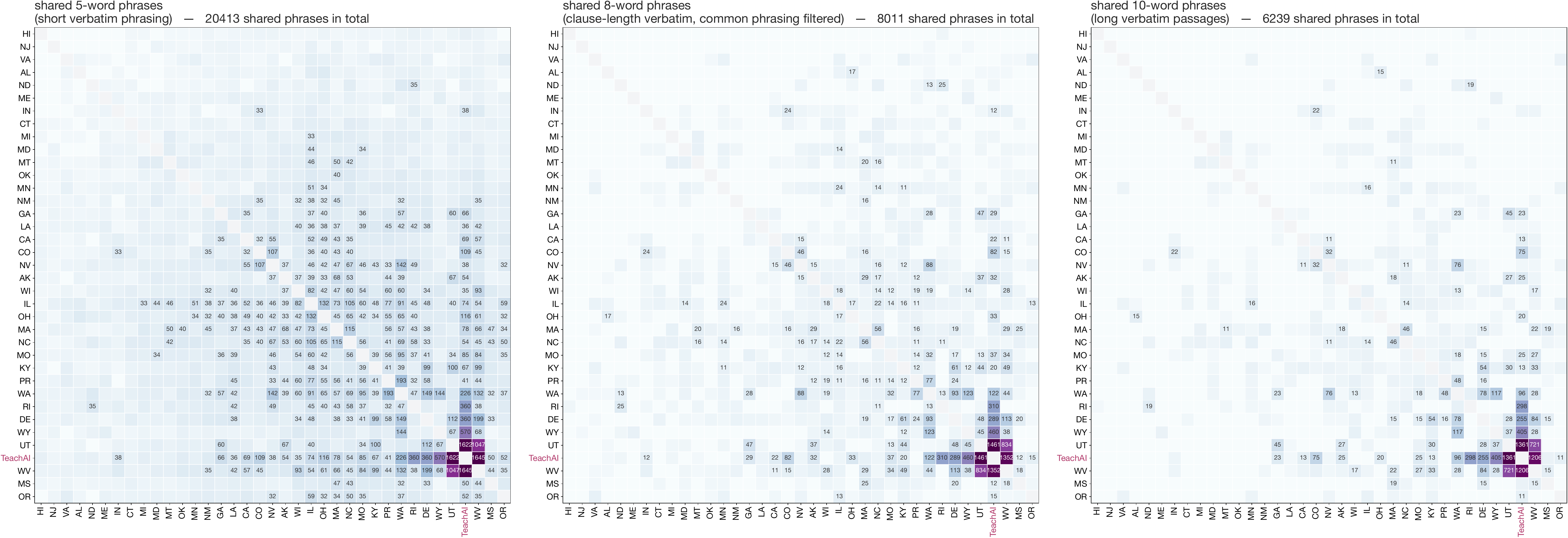}
  \caption{Verbatim phrasing shared between K-12 AI guidance documents. Each cell
  counts the distinct exact word $n$-grams two documents share (overlapping
  word-shingles), at rising phrase lengths: 5, 8, and 10-word.
  Darker = more shared verbatim text; undirected (no direction of influence implied).
  Documents ordered by hierarchical clustering on the 5-gram matrix (same order in all panels).}
  \label{fig:ngram-copying}
\end{figure*}

\begin{figure*}[!ht]
  \centering
  \includegraphics[width=\textwidth]{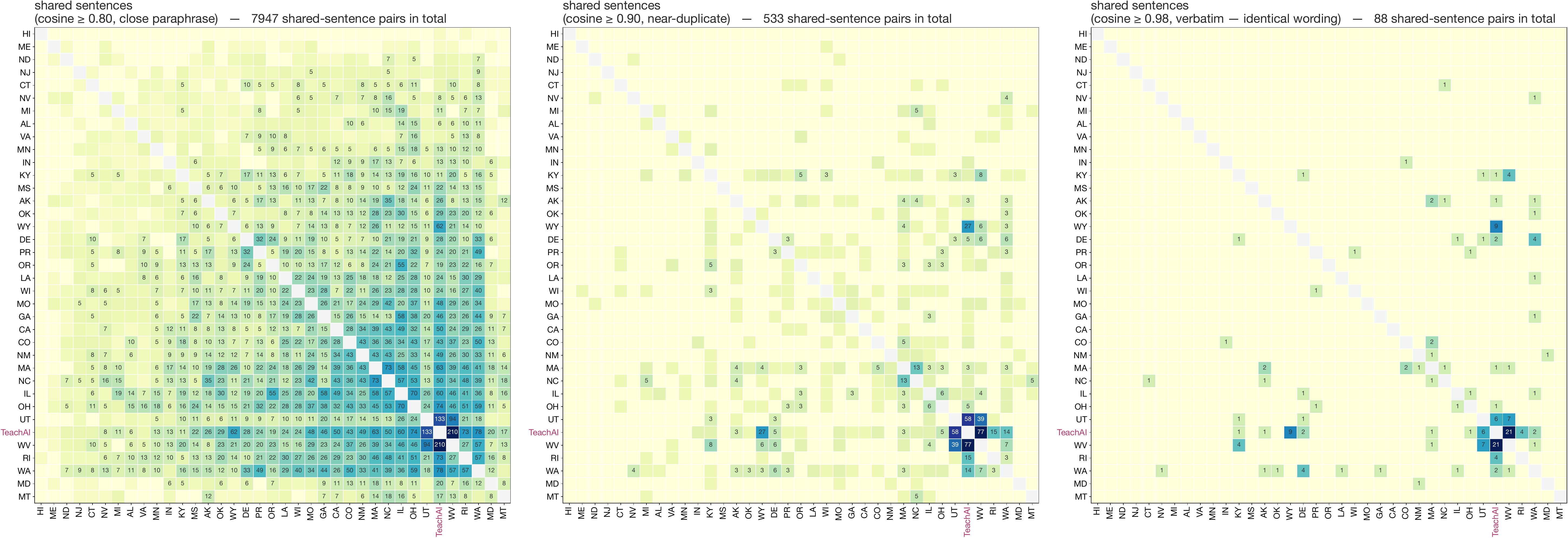}
  \caption{Semantically near-duplicate sentences shared between K-12 AI guidance
  documents (Sentence-BERT cosine). Each cell counts shared sentences at rising similarity
  thresholds: $\geq$0.80 (close paraphrase), $\geq$0.90 (near-duplicate), and
  $\geq$0.98 (verbatim). Darker = more shared sentences; undirected.
  Documents ordered by hierarchical clustering (same order in all panels).}
  \label{fig:cosine-sim}
\end{figure*}

\end{document}